\documentclass[journal]{IEEEtran}
\IEEEoverridecommandlockouts
\newenvironment{sequation}{\small\begin{equation}}{\end{equation}}

\pdfoutput=1
\usepackage{cite}

\usepackage{amsmath,amssymb,amsfonts}
\usepackage{algorithmic}
\usepackage{array, tabularx, makecell, colortbl}
\usepackage{diagbox}
\usepackage{booktabs}
\usepackage{graphicx}
\graphicspath{{figure/}}
\usepackage{algorithm,algorithmic}
\usepackage{lineno}

\usepackage{multirow}
\usepackage{textcomp}
\usepackage{xcolor}
\def\BibTeX{{\rm B\kern-.05em{\sc i\kern-.025em b}\kern-.08em
    T\kern-.1667em\lower.7ex\hbox{E}\kern-.125emX}}
\begin{document}

\title{STN: a new tensor network method to identify stimulus category from brain activity pattern}
%
%


\author{Chunyu Liu $^{1}$,\and Bokai Cao$^{2}$, \and Jiacai Zhang$^{3\ast}$

%

\thanks{
This work is funded by the General Program (61977010) of the Nature Science Foundation of China to J.Z. and Project funded by China Postdoctoral Science Foundation (2022M710210) to C. L. The authors would also like to thank all anonymous participants of the MEG experiments.

Chunyu Liu is with the school of Psychological and Cognitive Science and Beijing Key Laboratory of Behavior and Mental Health, Peking University, Beijing 100871,China(e-mail:lcy127@pku.edu.cn).
Bokai Cao is currently a research scientist at Meta Inc.
Jiacai Zhang is with School of Artificial Intelligence, Beijing Normal University, Beijing 100875, China(e-mail:jiacai.zhang@bnu.edu.cn).
}
}

\maketitle
\begin{abstract}

Neural decoding is still a challenge and hot topic in neurocomputing science. Recently, many studies have shown that brain network patterns containing rich spatial and temporal structure information, which represents the activation information of brain under external stimuli. 
The traditional method extracts brain network features directly from the common machine learning method, then puts these features into the classifier, and realizes to decode external stimuli. However, this method cannot effectively extract the multi-dimensional structural information, which is hidden in the brain network. The tensor researchers show that the tensor decomposition model can fully mine unique spatio-temporal structure characteristics in multi-dimensional structure data. This research proposed a stimulus constrained tensor brain model(STN)which involves the tensor decomposition idea and stimulus category constraint information. The model was verified on the real neuroimaging data sets (MEG and fMRI). The experimental results show that the STN model achieves more than $11.06\%$ and $18.46\%$ on accuracy matrix compared with others methods on two modal data sets. These results imply the superiority of extracting discriminative characteristics about STN model, especially for decoding object stimuli with semantic information.

\end{abstract}



\section{Introduction}

\IEEEPARstart{T}{he} process of neural decoding is through analyzing the neural signals pattern, which was collected by the non-invasive device, to analyze the neural activation pattern in response to specific visual stimuli, then using this neural activation pattern to deduce the external stimulus categories inversely~\cite{kriegeskorte2019interpreting}. Where one of the vital steps is to establish the mapping relationships among the neural activation pattern, neural signals pattern, and external stimuli. Previous research mainly focused on the brain individual area model to establish this mapping relation(such as single brain area or channel signals)~\cite{2011Decoding}. But now, it become  hot research to study the mapping from the network model on the level of the whole brain or the local system~\cite{2020Decoding,Anzellotti2018Beyond}.

Recently, the methods of brain network patterns have been successfully applied to decode the cognitive state, such as emotional category\cite{fatemeh2019statistical},object category~\cite{wang2019decoding}, visual attention~\cite{parhizi2018decoding}, and tracking the ongoing cognitive state~\cite{gonzalezcastillo2015tracking}, etc.
The brain network structure data is usually present involving dimensions, such as time and space dimensions, and indirectly representing potential characteristics, such as visual stimuli and participant characteristics~\cite{cao2015a}. Indeed, some researchers also show that this brain connection patterns not only has a spatiotemporal structure, but is also as the carrier between neuroimaging data and stimulus information\cite{Anzellotti2018Beyond}. 
However, the traditional data processing method  directly inputs the brain network structure into the classifier, firstly is to extract the brain network features and then sends it to the classifier for classification, that lead to losing some multi-dimensional structural characteristics of the brain network. Therefore, it is necessary to build a stronger representation ability model to extract network features.

Tensor data is a kind of structured data, which can retain the high-order statistical characteristics of original data, internal structure information, and the correlation between data dimensions. The brain network data is a kind of small sample and high dimensional data, which own the characteristics of spatiotemporal structure. In order to improve the ability of mining brain network features,  the expression form of tensor data is introduced to represent the brain network data, then building a decoding model through combining the mature tensor model theory and some prior knowledge of brain science that will significantly improve the decoding effect of brain activity.

In the field of neural signals, some researchers have introduced tensor models and network features into neuroimaging data analysis, such as EEG data\cite{huang2019tensor} and fMRI data\cite{2017Detecting}.
For example, the research directly uses the tensor structure model to analyze the dynamic changes in the  brain network\cite{ZHU2020116924} and detect the topological structure and other characteristics\cite{alsharoa2019tensor,Topology2020}.

There is a part of the research introduce multi-view information-constrained tensor network to extract the brain network pattern characteristics from EEG\cite{Cao2017T-BNE} or fMRI data\cite{alsharoa2019tensor,Topology2020}, then to distinguish between normal and patients. At present, some results of brain network model show that the effect of tensor-based model is better than that of the matrix-based model\cite{2018Tensor}. However, there are still some shortcomings in the brain network model based on tensor form, such as ignoring the sparsity in the network, local optimization, and some prior information about the constraint relationship between stimulus features and networks\cite{Topology2020}.

Among the previous methods of neural decoding, there is a type of research using visual stimulus information to build the neural coding and decoding model. It introduced visual stimulus information into the model training process,  using the characteristic neural data to optimize the parameters of classifier under the aid or constraint of the stimulus information~\cite{cowen2014neural}. This type of model considers the prior information of visual stimuli and the coupling information pattern between brain signal, which makes the effect of the decoding model is much higher than other models. Inspired by this, this research attempted to construct a tensor brain network model based on visual stimulus constrains, which is called the Stimulus-constrained-Tensor brain Network (STN). While constructing the STN model, we introduced the idea of the compactness about the stimulus category according to the brain network pattern into the framework of multivariate analysis.

The main contents of this research are as follow: 1. constructing the Stimulus-constrained-Tensor brain Network model, 2. constructing brain network pattern of different visual stimuli under fMRI and MEG  modalities data sets, then adopting semi-supervised learning classification framework to verify the effective of model, 3. analysis and interpretation about parameters of model.

\section{Method}

\begin{figure*}
  \centering
  \includegraphics[width=0.95\linewidth]{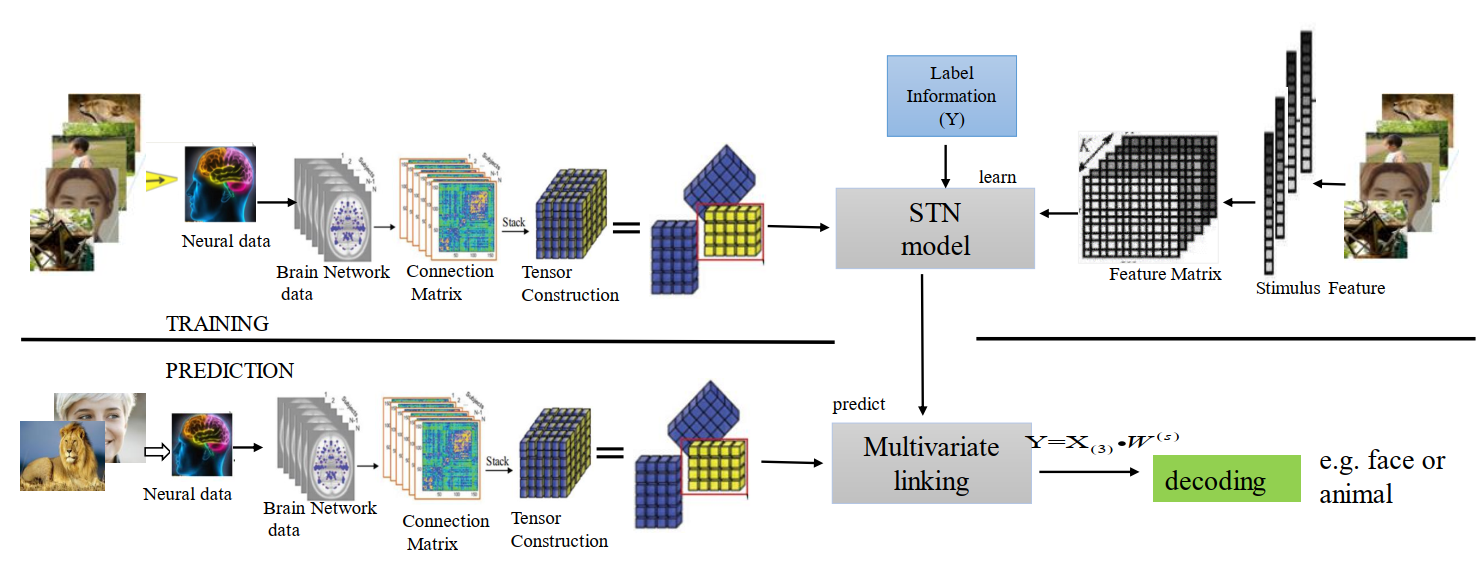}
  \caption{The schematic diagram of Stimulus-constrained-Tensor brain Network model}\label{fig:1.1}
\end{figure*}

This paper proposes a tensor brain network model based on visual stimuli constraints coupled with multivariate pattern analysis to realize decoding  visual stimulus categories.  In the section of Method, the paper will introduced four aspects. Firstly, we defined the data sets of brain network pattern under visual stimulation. Next, we introduce the tensor decomposition technology and build a tensor brain network based on stimulus constraints. Finally, we introduce the optimization algorithm to solve the model. The schematic diagram of the model framework is shown in the figure\ref{fig:1.1}.

\subsection{The definition of graph structure in brain network}

The paper used graph structure to represent the structure of brain network data. It is assumed that a type of neural data under visual stimuli could construct a set of corresponding brain network patterns. Firstly, setting $ \textstyle D=\{G_ {1},\ldots,G_{n}\}$ to represent a series of brain network pattern, where $n$ indicates the number of visual stimuli of the same kind.  The paper supposed that the brain network structure under visual stimulation share a series of vertex sets $\scriptstyle V$,  represents a specific brain anatomy template so that the brain can be divided into $m$ brain regions. For A brain network $ G_{i}$  can be represent by an adjacency matrix $ A_{i}\in \Re^{m\times m}$.

{\bfseries Definition 1} (Graph) The set $ G=(V,E)$ represents a graph, where, $ V=\{v_{1},\ldots,v_{m}\}$ represent a set of vertices, $ E\subseteq V\times V$ represents a set of edges.

In reality, it is a laborious task to label data, especially for this complex data, such as graph structures. Therefore, the paper adopts a semi-supervised learning framework to train brain network data; that is, part of the data is labeled, and some of the data is unlabeled. For example, during the brain network set of the $D$, the $lth$ sample is labeled with $  Y\in\Re^{l\times c}$, where $c$ is the number of class labels. Each brain network only belongs to one category, that is, if $ G_{i}$ belongs to $jth$ category, then  $ Y(i,j)=1$. To simplify the labeling, the paper will label the labeled data as $ D_ {l}=\{G_{1},\ldots,G_{l}\}$, the unlabeled data set as $D_{u}=\{G_{l+1},\ldots,G_{n}\}$, and the total data set as $ D=D_{l}\bigcup D_{u}$.

\subsection{ Tensor brain network decomposition}

 In this paper, the output form of graph structure is represented by a tensor, and the structure characteristic of the brain network is learned in the space represented by the tensor. First of all, the brain network data under visual stimulation, that is $\{A_{i}\}_{i=1}^{n}$, were combined into a large partial symmetric tensor structure $\mathcal{X}\in\Re^{m\times m\times n}$, then used tensor technology to model and analyze the tensor brain network.

{\bfseries Definition 2} (Partially symmetric tensor) A $m$ order of tensor $\mathcal{X}\in\Re^{I_{1}\times \cdots\times I_{m}}$ is a partially symmetric tensor, if, it can be represented by the tensor product of $m$ vectors on the mode of $i_ {1},\cdots,i_{j}\in\{1,\cdots,m\}$, that is
\begin{sequation}\label{eq:partensor}
 \mathcal{X}=x^{(1)}\circ\cdots\circ x^{(m)}
\end{sequation}
Where, $x^{(i_{1})}= \cdots=x^{(i_{j})}$.

{\bfseries Definition 3} (CP decomposition) For any tensor $\mathcal{X}\in\Re^{I_{1}\times \cdots\times I_{m}}$, its
CP decomposition is to decompose the tensor into the sum of a number of rank, and its form is shown in formula \ref{eq:cptensor}.
\begin{sequation}\label{eq:cptensor}
 \mathcal{X}\approx \sum_{r=1}^{k}x_{r}^{(1)}\circ\cdots\circ x_{r}^{(m)}\equiv\mathcal{C}\times_{1}X^{(1)}\cdots\times_{M}X^{(m)}
\end{sequation}
Where, $\mathcal{C}\in\Re^{k\times \cdots\times k}$ is indicative tensor, that is $\mathcal{C}\{i_ {1},\cdots,i_{k}\}=\delta(i_{1}=\cdots=i_{k})$, $k$ is the number of factors.

This research combined the output form of brain network data into a form with a tensor structure, and assumed this tensor structure involved three dimensions of information, such as time, space, and visual stimuli. Since the construction process of brain network data was designed to time and space information, and in the graph structure, the spatio-temporal information is represented by the connection information between the vertices and the vertices. Therefore, this study writed the CP decomposition form of the third-order tensor brain network into the following formula \ref{eq:cpnetwork}.

\begin{sequation}\label{eq:cpnetwork}
 \mathcal{X}\approx \sum_{r=1}^{k}B_{:,r}^{(s)}\circ B_{:,r}^{(s)}\circ S_{:,r}=\mathcal{C}\times_{1}B^{(s)}\times_{2}B^{(s)}\times_{3}S
\end{sequation}

\begin{figure}[!ht]
  \centering
  \includegraphics[width=0.65\linewidth]{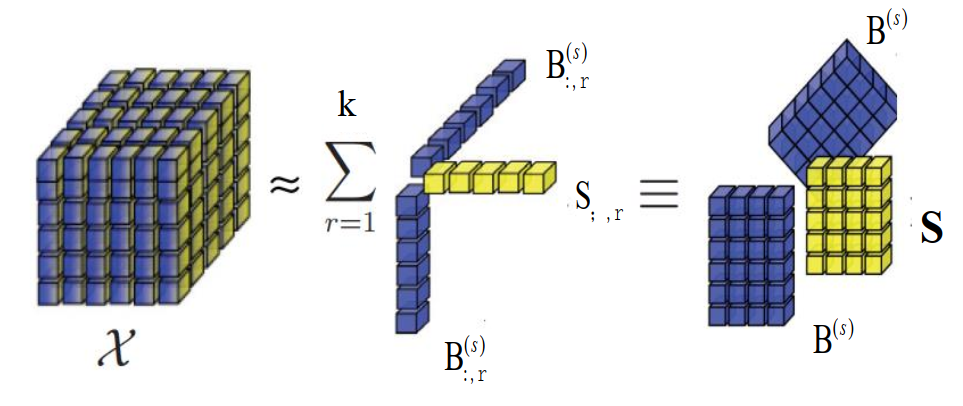}
  \caption{The diagram of CP decomposition about tensor brain network}\label{fig:5.3}
\end{figure}

Where $B^{(s)}\in \Re^{m\times k}$ is the vertex factor matrix, $S\in \Re^{n\times k}$ is the stimulus factor matrix, $\mathcal{C}\in\Re^{k\times k\times k}$ is the indicative tensor. The schematic diagram of CP decomposition is shown in figure \ref{fig:5.3}.

In fact, the more hidden information about the brain network structure can be extracted from the tensor coupling space, through embedding the tensor brain network into the learning process. For the above formula \ref{eq:cpnetwork}, it is to learn the representation information of brain network structure space, which constructed under the constraints of visual stimuli. The purpose of tensor brain network model is to find discriminative structural information, so that it is more easily to separate the brain network patterns with different labels. 

\subsection{The tensor brain network model based on stimulus constraints}

The following three problems need to be solved during constructing STN model: (1) how to learn the high-order structural characteristics of the brain network during the process of representation learning, (2) how to add visual stimulus information as an auxiliary variable to learn the brain network data, (3) how to integrate the learning of the classifier into the process of representation learning. For the first problem, this paper has been solved in constructing tensor brain network data. Next, the paper will focus on how to solve the latter two problems.

The stimulus information can be as a feature to assist in the analysis of the brain signal data. Previous studies have pointed out that in the hidden space of auxiliary variables\cite{narita2011Tensor}, the similarity of the brain network structure under visual stimulus is consistent with the similarity of the stimulus feature space. When the two visual stimulus features are close, the characteristics extracted from the brain network structure are also close. Therefore, this paper defined an objective function to constrain the distance of brain network structure based on visual stimulus characteristics.

\begin{sequation}\label{eq:sideinformation}
\min_{S}\sum_{i,j=1}^{n}\|S(i,:)-S(j,:)\|^{2}_{F}Z^{(s)}(i,j)
\end{sequation}
Where $Z$ is the kernel matrix, its whole $Z^{(s)}(i,j)$ represents the similarity of the network structure in the stimulus feature space. In this case, the boundary information about the visual stimulus features can be effectively used as a guide to discover meaning hidden factors. For simplicity, the formula \ref{eq:sideinformation} is rewritten as

\begin{sequation}\label{eq:sidechong}
\min_{S} tr(S^{T}L_{Z^{(s)}}S)
\end{sequation}

Where the symbol $tr(\cdot)$ represents the trace of a matrix, $L_{Z^{(s)}}$ is the Laplacian matrix which is derived from the similarity matrix $Z$, that is ${Z^{(s)}}=D_{Z^{(s)}}-Z$, and $D_{Z^{(s)}}$ is the diagonal matrix, its non-zero elements are the sum of the row vectors of the matrix, specifically is $D_{Z^{(s)}}(i,i)=\sum_{j}Z^{(s)}(i,j)$.

In addition, For the brain network model under the constraint of stimulus features, in order to better discover hidden discriminative factors and obtain higher accuracy and interpretable results, this paper adopts orthogonal constraints on the stimulus matrix factors.

\begin{sequation}\label{eq:sideyueshu}
S^{T}S=I
\end{sequation}

Aiming at the third question, that is, to learn and classify brain network structure data. This paper supposed there was a mapping matrix $W^{(s)}\in \Re^{k\times c}$ between the visual stimulus $S$ and the label $Y$. So the third question can be described by the ridge regression problem as
\begin{sequation}\label{eq:classridge}
\min_{W^{(s)}}\|DSW^{(s)}-Y\|_{F}^{2}+\gamma\|W^{(s)}\|_{F}^{2}
\end{sequation}

Where, $D=[I^{l\times l}$, $0^{l\times (n-l)}]\in \Re^{l\times n}$, $\|W^{(s)}\|_{F}^{2}$ control the capacity of $W^{(s)}$, the parameter $W^{(s)}$ control the influence of $W^ {(s)}$.

This research adopts the hidden factor $S$ on the stimulus feature modality as a feature, combining the framework of semi-supervised learning to find the discriminative features related to classification from original data. In the classification learning framework, this paper embed the hidden feature $S$ and the parameter of classification learning framework $W^{(s)}$ into the same model so that they can be cross-trained with each other. Where the hidden feature $S$ is learned from the tensor decomposition model. In this case, combined with partial symmetry tensor and tensor decomposition technology, this paper indirectly put brain network data $\mathcal{X}$ and visual stimulus category information $Y$ coupling into the visual stimulus modal space for training and learning.

Finally, the STN model proposed in this study can be described by the following optimization problem.
\begin{sequation}
\label{eq:mainfunction}
\begin{aligned}
& \min_{B^{(s)},S,W^{(s)}}\|\mathcal{X}-\mathcal{C}\times_{1}B^{(s)}\times_{2}B^{(s)}\times_{3}S\|_{F}^{2}+\alpha tr(S^{T}L_{Z}S)\\
&\quad \quad +\beta\|DSW^{(s)}-Y\|_{F}^{2}+\gamma\|W^{(s)}\|_{F}^{2} \\
& s_{\cdot}t_{\cdot}\quad  S^{T}S=I
\end{aligned}
\end{sequation}

Where, $\alpha$,$\beta$,$\gamma$ separately control the three parameter values of stimulus information, classification loss and regularization.

\subsection{The process of solving}

The process of solving the STN model is mainly to estimate the parameters $B^{(s)}\in \Re^{m\times k}$, $S\in \Re^{n\times k}$  and $W^{(s)}\in \Re^{k\times c}$. Since the formula \ref{eq:mainfunction} is a non-convex orthogonal constrained optimization problem, it isn't easy to find the optimal global solution of the model. In order to solve the above problems, this paper adopt a framework of solving parameter alternately, that is, to transform the above optimization problem into fixing other parameters first, and solving a parameters optimization problem, iterating the parameters until the model converges. This study used the Alternating Direction Method of Multipliers~\cite{Eckstein1992On} to solve the STN model.

\noindent{\bfseries Fixing the parameters $S$ and $W^{(s)}$, solving $B^{(s)}$}

$\mathcal{X}$ is a partially symmetric tensor, and the formula \ref{eq:mainfunction} involve the fourth-order term of $B$, so it is difficult to directly solve the optimal solution. This paper introduced the variable substitution technology. The original function was transformed into minimizing the formula \ref{eq:Bmin}.
\begin{sequation}
\label{eq:Bmin}
\begin{aligned}
&\min_{B^{(s)},F^{(s)}}\|\mathcal{X}-\mathcal{C}\times_{1}B^{(s)}\times_{2}F^{(s)}\times_{3}S\|_{F}^{2}\\
&   s_{\cdot} t_{\cdot} \quad  F^{(s)}=B^{(s)}
\end{aligned}
\end{sequation}

Where, $F^{(s)}$ is an auxiliary variable. The augmented lagrangian function of formula \ref{eq:Bmin} could be written as
\begin{sequation}
\label{eq:augmented}
\begin{aligned}
&\mathcal{L}(B^{(s)},F^{(s)},U,\mu)\\
&=\|\mathcal{X}-\mathcal{C}\times_{1}B^{(s)}\times_{2}F^{(s)}\times_{3}S\|_{F}^{2}+\frac{\mu}{2}\|F^{(s)}-B^{(s)}\|_{F}^{2}\\
& +tr(U^{T}(F^{(s)}-B^{(s)}))
\end{aligned}
\end{sequation}

Where, $U\in\Re^{m\times k}$ is Lagrangian multiplier, $\mu$  is penalty multiplier, the parameter of $\mu$ is adjusted according to the conference\cite{lin2011Linearized}. In order to find the optimal solution of B, this paper rewrote the formula \ref{eq:augmented} into the form of a convex function. The specific calculation process is as follows, firstly, during the tensor decomposition, the tensor can be written an equivalent form according to modal one:
\begin{sequation}\label{eq:tensormode}
X_{(1)}\approx B^{(s)}(S\odot F^{(s)})^{T}
\end{sequation}
Then, the tensor decomposition form in formula \ref{eq:Bmin} can be further written as

\begin{sequation}\label{eq:tensormode1}
\begin{aligned}
\|\mathcal{X}-\mathcal{C}\times_{1}B^{(s)}\times_{2}F^{(s)}\times_{3}S\|_{F}^{2}
=\|B^{(s)}E^{T}-X_{(1)}\|_{F}^{2}
\end{aligned}
\end{sequation}

Where $E=S\odot F^{(s)}\in \Re^{(m\ast n)\times k}$, $\odot$ represents the Khatri-Rao product.

For the formula \ref{eq:augmented}, we added one $\frac{1}{\mu}\|U\|^{2}$ to the last two items, it becomed
\begin{sequation}\label{eq:tensor1}
\begin{aligned}
& tr(U^{T}(F^{(s)}-B^{(s)}))+\frac{\mu}{2}\|F^{(s)}-B^{(s)}\|_{F}^{2}\\
&=\frac{\mu}{2}\|B^{(s)}-F^{(s)}-\frac{1}{\mu}U\|_{F}^{2}-\frac{1}{2\mu}\|U\|^{2}_{F}
\end{aligned}
\end{sequation}
Since the last item in formula \ref{eq:tensor1} has irrelevant to $B^{(s)}$, combined with formula \ref{eq:tensormode1}, the minimum value of the parameter $B^{(s)}$  can be transformed into an augmented lagrangian function, that is
\begin{sequation}\label{eq:tensorB}
\tilde{\mathcal{L}}(B^{(s)})=\min_{B^{(s)}}\|B^{(s)}E^{T}-X_{(1)}\|_{F}^{2} +\frac{\mu}{2}\|B^{(s)}-F^{(s)}-\frac{1}{\mu}U\|_{F}^{2}
\end{sequation}

In this way, the problem is simplified to a problem of the optimal solution $B^{(s)}$ of a convex function, and its optimal value is at the inflection point of the function. The specific solution derivation process is as follows:
\begin{sequation}\label{eq:tuidao1}
\begin{aligned}
\frac{\partial\tilde{\mathcal{L}}(B^{(s)})}{\partial B^{(s)}}
=2B^{(s)}E^{T}E-2X_{(1)}E+\mu B^{(s)}-\mu F^{(s)}-U=0
\end{aligned}
\end{sequation}
Combined $B$ item in formula \ref{eq:tuidao1}, and moved the unrelated items to the right of the equal, that is
\begin{sequation}\label{eq:tuidao2}
B^{(s)}(2E^{T}E+\mu I)=2X_{(1)}E+\mu F^{(s)}+U
\end{sequation}
Finally, the optimal solution  $B^{(s)}$ obtain in this paper is
\begin{sequation}\label{eq:tuidao3}
B^{(s)}=(2X_{(1)}E+\mu F^{(s)}+U)(2E^{T}E+\mu I)^{-1}
\end{sequation}
At the same time, the optimal solution of another auxiliary variable $F^{(s)}$ is the formula \ref{eq:tuidao4}.
\begin{sequation}\label{eq:tuidao4}
F^{(s)}=(2X_{(2)P}+\mu B^{(s)}-U)(2P^{T}P+\mu I)^{-1}
\end{sequation}
Where, $P=S\odot F^{(s)}\in \Re^{(m\ast n)\times k}$, $X_{(2)}$ is the matrixization of tensor $\mathcal{X}$ on mode two. In addition, this paper adopt the principle of gradient descent to optimize the lagrangian factor. The specific process is shown in formula \ref{eq:optlag}.

\begin{sequation}\label{eq:optlag}
U\longleftarrow U+\mu(F^{(s)}-B^{(s)})
\end{sequation}

\noindent{\bfseries Fixing the parameters $W^{(s)}$ and $B^{(s)}$, solving$S$}

The parameter $S$ is the factorization matrix of the tensor's model three on visual stimulus characteristic direction. So, we set the objective optimization function as follows:
 \begin{sequation}\label{eq:tuidao11}
\begin{aligned}
&\mathcal{L}(S)=\|SG^{T}-X_{(3)}\|_{F}^{2}+\alpha tr(S^{T}L_{Z}S)+\beta \|DSW^{(s)}-Y\|_{F}^{2}\\
&s_{\cdot}t_{\cdot} S^{T}S=I
\end{aligned}
\end{sequation}
Where, $G=F^{(s)}\odot B^{(s)}\in \Re^{(m\ast m)\times k}$, $X_{(3)}\in \Re^{n\times(m\ast m)}$ is the matrix on tensor mode three. The objective function is an optimization problem with orthogonal constraints.

At present, many researchers have proposed algorithms to solve this orthogonal constraint problem\cite{gao2019parallelizable,OptimizationAlgorithmsonMatrixManifolds}. In this research, this article used the common method, which is the bilinear search method to solve the problem\cite{10.1007/s10107-012-0584-1}. Firstly, we calculated the derivative of $\mathcal{L}(S)$ respect to $S$.
\begin{sequation}\label{eq:optlag1}
\nabla_{S}\mathcal{L}(S)=SG^{T}G-X_{(3)}G+\alpha L_{Z}S+\beta D^{T}(DSW^{(s)}-Y)(W^{(s)})^{T}
\end{sequation}

\noindent{\bfseries Fixing the parameters $S$ and $B^{(s)}$, solving$W^{(s)}$}

The last part is the process of solving the parameter $W^{(s)}$, which represent the process of classification learning. From the objective function formula \ref{eq:optw}, the main function is the least square function under the quadratic constraint, and it is a convex function, so there is an optimal solution.
\begin{sequation}\label{eq:optw}
\mathcal{L}(W^{(s)})=\|DSW^{(s)}-Y\|_{F}^{2}+\gamma\|W^{(s)}\|_{F}^{2}
\end{sequation}
The process of solving this parameter is that the objective is zero after deriving $W^{(s)}$, the inflection point is the optimal solution $W^{(s)}$. The specific process is as follows:


\begin{sequation}\label{eq:wtuidao1}
\begin{aligned}
\frac{\partial\mathcal{L}(W^{(s)})}{\partial W^{(s)}}
=2(S^{T}D^{T}DS+\gamma I)W^{(s)}-2S^{T}D^{T}Y=0
\end{aligned}
\end{sequation}

Shift the terms of formula \ref{eq:wtuidao1}, it can be simplified as
\begin{sequation}\label{eq:wtuidao2}
(S^{T}D^{T}DS+\gamma I)W^{(s)}=S^{T}D^{T}Y
\end{sequation}
The finally to solved $W^{(s)}$ is the formula \ref{eq:wtuidao3}.
\begin{sequation}\label{eq:wtuidao3}
W^{(s)}=(S^{T}D^{T}DS+\gamma I)^{-1}S^{T}D^{T}Y
\end{sequation}
Last, the solution process of the tensor brain network based on visual stimulus constraints is shown in Algorithm 1.

\begin{algorithm}[h]
  \renewcommand{\algorithmicrequire}{\textbf{Input:}}
  \renewcommand{\algorithmicensure}{\textbf{Output:}}
  \caption{STN}
  \label{alg:suanfa}
  \begin{algorithmic}[1]
   \REQUIRE $\mathcal{X}$,$\mathbf{Z,Y}$,$\alpha,\beta,\gamma$
   \ENSURE $\mathbf{B^{(s)},S,W^{(s)}}$
   \STATE set $\mu_{max}=10^{4},\rho=1.0$
   \STATE initialize:$\mathbf{B^{(s)},S,W^{(s)}}\sim \mathcal{N}(0,1),\mathbf{U}=0,\mu=10^{-4}$
   \REPEAT
   \STATE Update $B^{s}$ and $F^{s}$ by Eq.(\ref{eq:tuidao3})and Eq.(\ref{eq:tuidao4})
   \STATE Update $U$ by Eq.(\ref{eq:optlag})
   \STATE Update $\mu$ by $\mu\leftarrow min(\rho\mu,\mu_{max})$
   \STATE Update $S$ by Eq.(\ref{eq:optlag1})with curvilinear search
   \STATE Update $W^{(s)}$ by Eq.(\ref{eq:wtuidao3})
   \UNTIL{convergence}
 \end{algorithmic}
\end{algorithm}

\section{Experiments}

To test the effect of the model, this paper adopts two real collected data sets,  MEG data, which collected about four types of object pictures, and fMRI data, which collected about four kinds of emotional face pictures. At present, these two modals of neuroimaging data sets have been proved by many studies that they can construct the stable brain network pattern\cite{2019Network,ajmera2020decoding}.

\subsection{fMRI data}

Six Chinese participants were included the study (3 females, mean 24$\pm$2 years). All participants were right-handed, had normal hearing and had normal or corrected to normal vision. All participants successfully participated in an emotional stimulus picture test system, indicating that the participants have the ability to recognize emotions. In addition, all participants provided written informed consent. The Beijing normal University Review Board approved the experimental procedures.

A 3-T Siemens scanner equipped for echo planar imaging (EPI) was used for image acquisition. The functional images were collected with the following parameters: repeat time (TR)= 200ms; echo time (TE)= 30ms; 33 slices, matrix size = $64\times64$;acquisition voxel size = $3.125\times3.125\times4.2$; flip angle (FA)=$90^{\circ}$; field of view (FOV)= 200 mm.

Stimuli came from the NimStim set of facial expressions\cite{tottenham2009the}, the California Facial expressions of Emotion data set\cite{dailey2001california}, the Japanse Female Facial Expression Database\cite{lyons1998coding}, the Karolinska Directed Emotional Faces\cite{dailey2001california}, the Radboud Faces Database collection\cite{langner2010presentation}, 200 actors (105 males)portrayed each of four expressions: fear, disgust, happiness and neutral.

Participants performed two sessions, and each session included 5 fMRI runs. Each run started and ended with an 8s fixation baseline period. For each run, the expressions of 20 actors were shown. Each stimulus was presented for 2000ms followed by a 6000ms blank screen. Each 2000ms presentation consisted of a stimulus being flashed ON-OFF-ON-OFF-ON, where ON corresponds to presentation of the stimulus for 200ms, and OFF corresponds to presentation of the gray background for 200ms. A green fixation was shown on the screen during the whole experiment. Subjects were instructed to keep their eyes on the fixation and press different buttons with the right thumb to indicate the type of expressions.

\begin{figure}[!ht]
  \centering
  \includegraphics[width=0.75\linewidth]{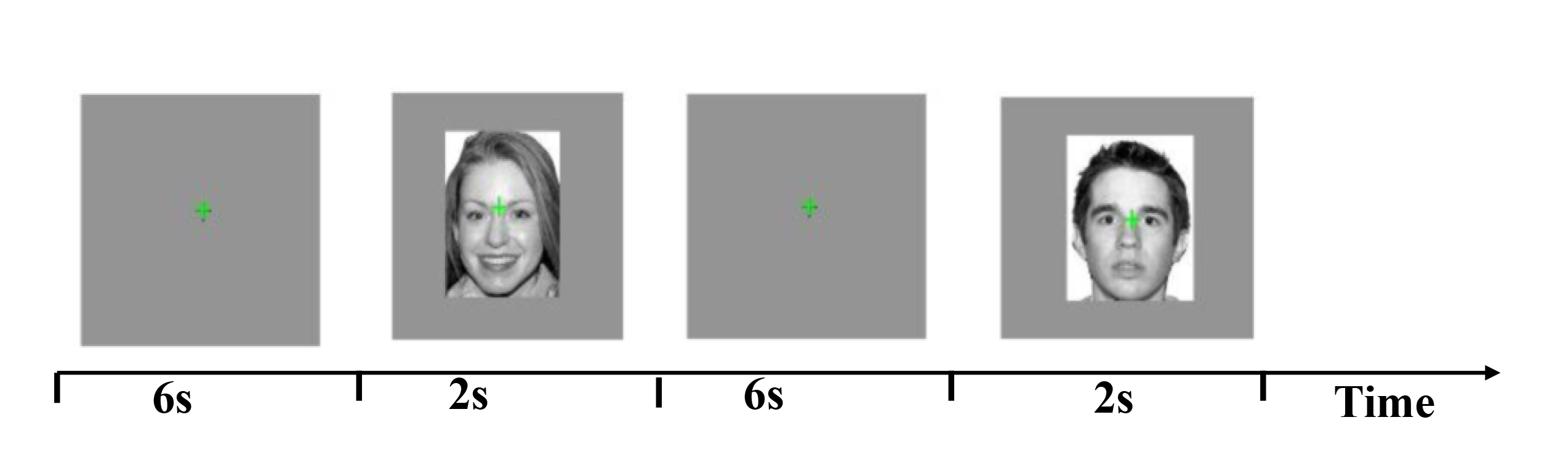}
  \caption{The experimental paradigm}\label{fig:5.2}
\end{figure}

\subsection{MEG data}

Nineteen Chinese participants were included the study (10 females, mean $23 \pm 2 $years). All participants were right-handed, had normal hearing and normal or corrected-to-normal vision. All participants provided written informed consent. Experimental procedures were approved by the Peking University Institutional Review Board.

The stimuli comprised  640 gray-scale images ($300 \times 300$ pixels, visual angle $7.92^{\circ}\times7.92^{\circ} $)from four categories (160 images per category), including faces with neutral expressions, scenes, animals and tools. All images were matched for mean luminance and contrast using the SHINE toolbox. The face stimuli, selected from the Chinese affective picture systems, include 80 unique male and female neural faces. The stimuli were all outdoor scenes, including mountains, countryside scenes, streets, and buildings, with 40 unique pictures of each type. The animal stimuli included mammals, birds, insects, and reptiles, composed of 40 items, and each with four exemplars. Finally, the tool stimuli included kitchen utensils, farm implements, and other common indoor tools, also comprising 40 items, each with four exemplars.

The experiment consisted of 10 runs. During each run, 64 visual stimulus images from four categories (16 faces, 16 scenes, 16 animals, and 16 tools)were presented to the participants randomly. Each image presentation lasted for 2000ms and was followed by a blank screen with the inter-stimulus intervals ranging randomly from 1500 ms to 2000ms. In each run, participants were instructed to press a button with their right index finger if the stimuli were the same. In addition, the participants were asked to concentrate on a fixed point (a green cross)in the center of the white screen.

\begin{figure}[!ht]
  \centering
  \includegraphics[width=0.75\linewidth]{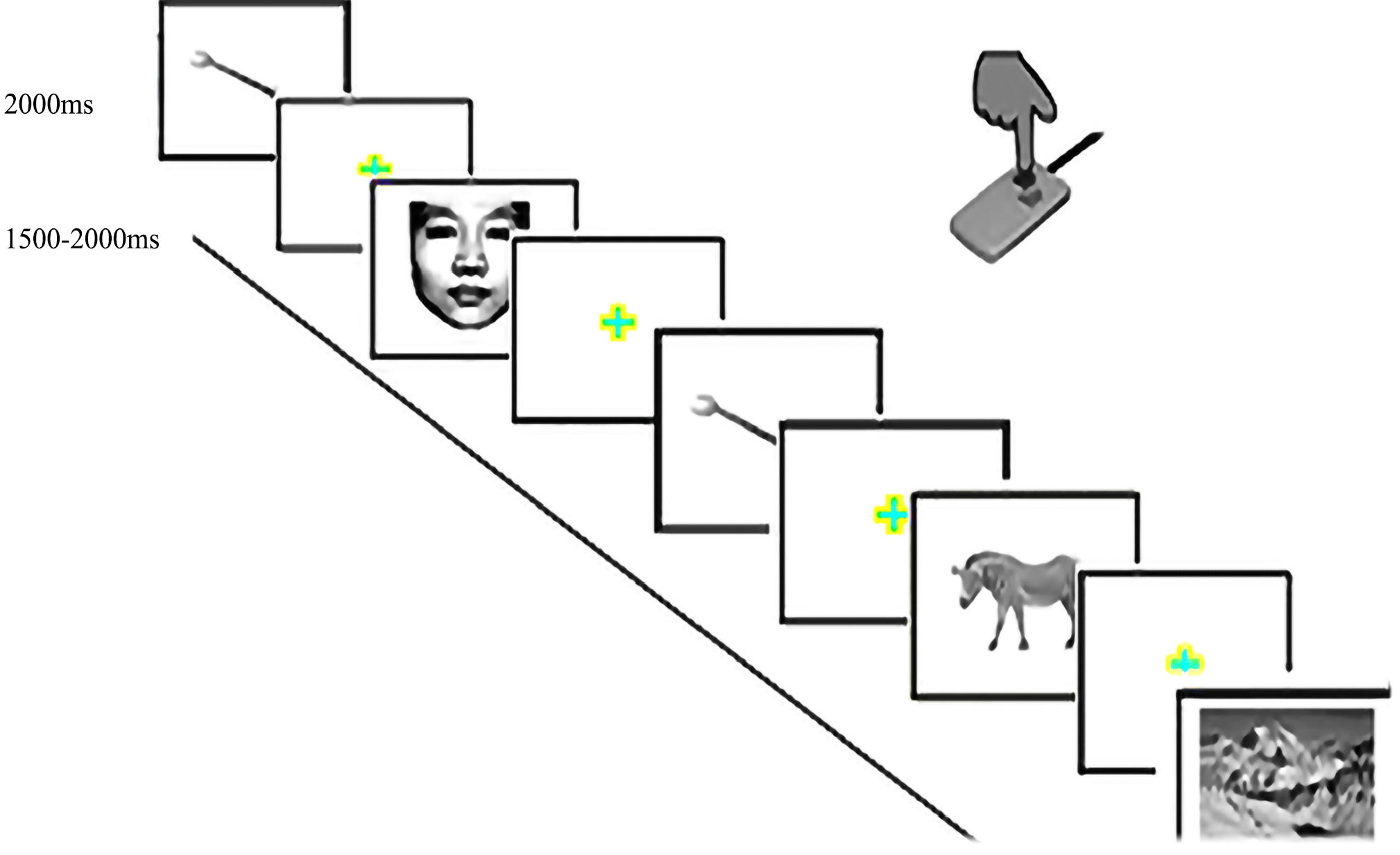}
  \caption{Schematic of the stimuli-viewing task}\label{fig:4.1}
\end{figure}

\section{Results}

 In the result, this paper first gave the results of the STN model on multiple data sets, and the decoding effect about other decoding models on the same data set. Then, we analyzed the effect of hyperparameters on the decoding results, which were involved in the STN model.

\subsection{The effect of model}

 In order to verify the effect of decoding the categories of visual stimulus in the STN model, this paper divided the two modalities' data into 14 data sets and then conducted experiments on these data sets one by one. Firstly, MEG data sets were divided into seven MEG data sets, including faces vs. animals, faces vs. scenes, faces vs. tools, animals vs. scenes, animals vs. tools, scenes vs. tools, and four types of visual stimulus data sets. fMRI data sets were divide into seven fMRI data sets, including disgust. vs neutral, disgust vs happiness, disgust vs neutral, surprise vs neutral, happiness vs neutral, and four emotion sets. Next, this article compared the decoding effect about three types of decoding models, and one was the traditional classification model based on brain network features, one was classification model based on the tensor brain network, and the last was the classification model based on Neural Network. Finally, this paper used three evaluation indicators to measure the model performance, including accuracy, recall, and F1 scores.

 \subsubsection{The results of decoding MEG data}

For constructing a brain network pattern in MEG data, we adopt the results of the references\cite{liu2020rapidly} to construct the optimal brain network pattern under four visual stimuli. The decoding results were the average of all subjects,  and each subjects' results were obtained through the average of the ten-fold cross-validation. At the same time, the decoding results of its various models on MEG data are presented in Table\ref{tab:4}. The values in parentheses represent the variance value between the subjects.

\begin{table*}[htbp]
  \centering
  \caption{Average decoding results of STN model and comparison model on MEG data}
  \label{tab:4}
  \begin{tabular}{c|c|ccccccc}
  \hline
  \multirow{2}{*}{Data}& \multirow{2}{*}{Measure}& &  & {Model} & &  &  & \\
  \cline{3-9}
   &  & SVM & ALS & Rubik & RF & LSTM & EEGNet & STN \\
  \hline
  \multirow{3}{*}{F$-$A}& ACC($\%$) & 84.21($\pm$4.21) & 71.44($\pm$2.32)& 57.44($\pm$2.32) & 85.31($\pm$2.12)& 64.79($\pm$3.42)& 89.11($\pm$3.12)& $\mathbf{95.10(\pm 2.52)}$\\
   & $F_{1}$ ($\%$) & 83.21($\pm$4.12) & 70.31($\pm $3.19)& 60.31($\pm$2.11) & 84.21($\pm$2.10)& 66.31($\pm$3.19)& 92.31($\pm$2.72)& $\mathbf{93.21(\pm2.52)}$ \\
   & Recall ($\%$) & 85.41($\pm$4.12) & 72.31($\pm $4.12)& 56.31($\pm$3.42) & 84.21($\pm$2.12)& 66.31($\pm$3.12)& 81.41($\pm$3.12)& $\mathbf{95.78(\pm2.12)}$\\
  \hline
  \multirow{3}{*}{F$-$S}& ACC($\%$) & 84.21($\pm$5.32) & 75.21($\pm$3.12)& 72.11($\pm$2.62) & 86.14($\pm$3.62)& 70.42($\pm$3.22)& 83.41($\pm$3.12)& $\mathbf{95.22(\pm 2.12)}$\\
   & $F_{1}$ ($\%$) & 85.31($\pm$4.32) &  73.39($\pm$2.17)& 70.20($\pm$4.02) & 88.19($\pm$3.02)& 71.14($\pm$2.12)& 84.24($\pm$3.62)& $\mathbf{95.31(\pm3.12)}$ \\
   & Recall ($\%$) & 86.14($\pm$3.52)& 74.39($\pm$2.12)& 73.29($\pm$1.12) & 87.12($\pm$2.12)& 70.42($\pm$2.12)& 83.19($\pm$3.12)& $\mathbf{94.34(\pm1.92)}$\\
  \hline
  \multirow{3}{*}{F$-$T}& ACC($\%$) & 83.21($\pm$5.31)& 80.14($\pm$3.52)& 60.21($\pm$2.19) & 83.41($\pm$3.12)& 62.31($\pm$2.12)& 92.25($\pm$3.13)& $\mathbf{96.79(\pm2.12)}$\\
   & $ F_{1}$ ($\%$) & 82.21($\pm$2.45) & 81.12($\pm$3.12)& 61.41($\pm$3.12) & 83.56($\pm$2.12)& 60.15($\pm$2.52)& 90.31($\pm$2.74)& $ \mathbf{96.14(\pm1.82)}$ \\
   & Recall ($\%$) & 83.43($\pm$3.14) & 80.42($\pm$2.46)& 59.41($\pm$2.82) & 81.31($\pm$3.12)& 61.41($\pm$2.19)& 92.14($\pm$2.12)& $\mathbf{95.89(\pm2.17)}$\\
  \hline
  \multirow{3}{*}{A$-$S}& ACC($\%$) & 77.41($\pm$4.32)& 65.42($\pm$2.22)& 65.42($\pm$3.19) & 70.32($\pm$2.19)& 63.21($\pm$2.12)& $\mathbf{86.16(\pm2.05)}$& 68.78($\pm$2.13)\\
   & $F_{1}$ ($\%$) & 75.42($\pm$4.12) & 66.62($\pm$2.04)& 67.14($\pm$3.01) & 70.41($\pm$2.42)& 62.31($\pm$1.89)& $\mathbf{85.29(\pm2.03)}$& 67.38($\pm$2.19) \\
   & Recall ($\%$) & 76.41($\pm$4.19) & 65.31($\pm$1.97)& 66.14 ($\pm$2.62)& 69.31($\pm$2.62)& 63.24($\pm$3.14)& $\mathbf{87.13(\pm2.32)}$& 66.31($\pm$1.92)\\
  \hline
  \multirow{3}{*}{A$-$T}& ACC($\%$) & 68.42($\pm$5.23) & 66.12($\pm$2.27)& 60.42($\pm$1.82) & 66.41($\pm$3.12)& 60.24($\pm$2.11)& 70.14($\pm$2.02)& $\mathbf{72.42(\pm5.12)}$\\
   & $F_{1}$ ($\%$) & 67.25($\pm$2.01) & 65.29($\pm$2.17)& 61.19($\pm$2.19) & 65.29($\pm$3.07)& 60.39($\pm$2.72)& 69.41($\pm$2.23)& $\mathbf{70.28(\pm2.12)}$ \\
   & Recall ($\%$) & 66.19($\pm$5.19) & 67.38($\pm$2.32)& 62.24($\pm$2.41) & 64.41($\pm$1.92)& 61.15($\pm$2.17)& 70.41($\pm$1.92)& $\mathbf{71.38(\pm2.14)}$\\
  \hline
  \multirow{3}{*}{S$-$T}& ACC($\%$) & 75.15($\pm$4.32)& 60.17($\pm$3.12)& 65.42($\pm$1.92) & 70.17($\pm$2.52)& 60.41($\pm$2.19)& $\mathbf{88.18(\pm3.92)}$& 74.51($\pm$2.12)\\
   & $F_{1}$ ($\%$) & 74.51($\pm$4.12) & 59.53($\pm$2.72)& 64.27($\pm$2.13) & 70.45($\pm$2.42)& 61.21($\pm$2.52)& $\mathbf{86.31(\pm2.14)}$& 75.15($\pm$2.13) \\
   & Recall ($\%$) & 74.15($\pm$4.12) & 61.25($\pm$2.33)& 67.13($\pm$2.11) & 73.10($\pm$2.41)& 64.32($\pm$2.52)& $\mathbf{82.24(\pm1.92)}$& 73.16($\pm$1.89)\\
  \hline
  \multirow{3}{*}{Four}& ACC($\%$) & 55.42($\pm$4.12) & 43.24($\pm$2.24)& 42.28($\pm$2.32) & 44.37($\pm$2.19)& 39.29($\pm$2.31)& 53.26($\pm$2.34)& $\mathbf{64.32(\pm2.61)}$\\
   & $F_{1}$ ($\%$) & 54.41($\pm$2.42) & 49.34($\pm$2.33)& 44.41($\pm$2.19) & 40.43($\pm$2.19)& 41.34($\pm$1.95)& 53.29($\pm$2.42)& $\mathbf65.45(\pm2.11)$ \\
   & Recall ($\%$) & 54.59($\pm$4.14)& 41.10($\pm$2.19)& 44.42($\pm$1.12) & 43.19($\pm$2.42)& 43.48($\pm$2.62)& 52.34($\pm$1.92)& $\mathbf{63.49(\pm2.13)}$ \\
  \hline
  \end{tabular}
\end{table*}

In order to better measure the decoding effect of the STN model, this paper adopts the widely used and high-efficiency classification models, such as SVM and RF model, to extract brain network features and identified the different brain network pattern. These decoding model have strong applicability and do not require too many samples, so they are often used in visual decoding and have achieved good results~\cite{rezvani2019,Yang2020The}.
In this paper, the kernel parameters in the SVM model are a linear kernel, the other parameters are default. In addition, these models were implemented with scikit-learn on Python 3.6.

Secondly, to prove that the STN model proposed in this article can better decode the visual categories, this paper also compared the decoding models, which were constructed based on the tensor, such as Alternating Least Squares (ALS)and Rubik. Where the ALS model is a tensor technology based on the alternating least squares algorithm\cite{sharan2017}. It can effectively and quickly extract the tensor decomposition components, and does not require too many data samples, so ALS is currently commonly used to extract tensor brain network features. Rubik model is a  tensor decomposition model which is proposed based on prior knowledge constraints. It can effectively overcome the effects of noise and data missing to extract tensor components, indeed, it also can quickly perform parallel computing on large-scale neuroimaging data, so Rubik model is one of the reference templates commonly used to construct the knowledge constraint tensor models~\cite{Yin2018Joint,Yin2019Learning}. The ALS and Rubik models were implemented on MATLAB, and their parameters were refer to the setting values in the original literature\cite{Comon2009tensor,Wang2015Rubik}.

Finally, the deep learning model is the latest method to extract brain network features. However,since the data in this article is real collected neuroimaging data and its scale is small, we adopt the little scale neural network models, such as the shallow layers of LSTM and EEGNet.

LSTM is a neural network model, which is based on the RNN model, it can fully mine the time series information and semantic information about neuroimaging data\cite{Hochreiter1997long}. This type of neural decoding model was often used in the field of neural decoding, which is constructed based on the combination of LSTM and brain network features. In this experiment, the implementation of LSTM was on the MATLAB platform, the number of hidden layers was set to 128, the activation function was the cross-quotient loss function, the learning rate was 0.001, the $decay_rate$ was $0.1$, and the rest of parameters were default.

EEGNet is a compact convolutional network model. During the process of constructing the EEGNet model, it just involves the frequency domain filtering, time domain filtering, and spatial filtering\cite{Lawhern2018EEG}, and its structure is simplified and the parameters of the model are few. Therefore, it is very suited for the small amount data. In this experiment, EEGNet was implemented with pytorch platform on Python 3.6; the selected activation function was the cross-entropy loss function, the learning rate was 0.001, the batch size was set to be 32, the number of frequency domain filter convolution was 32, the number of spatial filter convolution was 2, dropout was set to be 0.25, Epoch was set to be 50.

Seen from the results of Table \ref{tab:4}, the average two classification decoding accuracy of the STN model is $83.80\%$, and the decoding rate of  four categories is $64.32\%$. These two decoding rates are significantly higher than the decoding rates of the other seven models. The experimental results show that the tensor brain network model can effectively extract more discriminative brain network features when the model added stimulus feature constraints, thereby significantly improving the effect of neural decoding.

 \subsubsection{The results of decoding fMRI data}

During the study of fMRI data classification, which was induced by emotional face pictures, the process of constructing brain network patterns was implemented on the MATLAB platform.

Firstly, the length of the window to construct the brain network must be determined. Taking into account the time delay characteristics of fMRI data, this article adopts a total of 60s duration under 10 similar visual stimuli. Where the duration of a single fMRI experiment included  2 seconds of stimulus presentation plus 4 seconds of fMRI time series moving backward. Secondly, Combined with Destrieux whole brain template~\cite{Destrieux2010Automatic} to extract the signal value of the whole brain area. Finally, the brain network pattern was obtained by calculating the time domain correlation between brain areas. The entire implementation process was through the DPABI software\cite{yan2016dpabi}. For the verification of the STN model on fMRI data, this article also adopts the same compared method on MEG data.

From the results presented in Table \ref{tab1:5}, the STN model showed the best on three indicators compared with other methods. These results indicate that the way of adding prior information constrained stimulus features to construct the tensor model,  which can help the tensor model extract more distinguishing structural features, especially for advanced cognition, such as decoding emotion activity state, and the effect are rather obvious.

 \begin{table*}[htbp]
  \centering
  \caption{Average decoding results of STN model and comparison model on fMRI data}
  \label{tab1:5}
  \begin{tabular}{c|c|ccccccc}
  \hline
  \multirow{2}{*}{Data }& \multirow{2}{*}{Measure}& & & {Model} & &  &  & \\
  \cline{3-9}
   &  & SVM & ALS & Rubik & RF & LSTM & EEGNet & STN \\
  \hline
  \multirow{3}{*}{D$-$F}& ACC($\%$) & 55.51($\pm$5.12) & 71.14($\pm$5.42)& 71.42($\pm$4.92) & 60.24($\pm$4.12)& 63.42($\pm$5.32)& 65.41($\pm$5.19)& $\mathbf{77.25(\pm5.34)}$\\
   & $F_{1}$ ($\%$) & 56.24($\pm$4.92) & 70.37($\pm$5.43)& 71.27($\pm$5.27)& 63.36($\pm$4.12)& 65.27($\pm$5.12)& 64.17($\pm$5.12)& $\mathbf{78.36(\pm5.12)}$ \\
   & Recall ($\%$) & 54.39($\pm$5.12) & 69.46($\pm$4.89)& 72.61($\pm$5.33) & 61.23($\pm$5.21)& 61.32($\pm$5.41)& 65.14($\pm$5.42)& $\mathbf{75.32(\pm4.97)}$\\
  \hline
  \multirow{3}{*}{D$-$H}& ACC($\%$) & 70.14($\pm$5.32)& 75.24($\pm$4.82)& 62.14($\pm$5.17) & 70.42($\pm$5.19)& 60.41($\pm$5.32)& 56.35($\pm$4.95)& $\mathbf{81.26(\pm3.12)}$\\
   & $F_{1}$ ($\%$) & 71.35($\pm$4.92)& 73.18($\pm$4.65)& 61.28($\pm$5.43) & 71.32($\pm$5.72)& 61.42($\pm$5.43)& 56.41($\pm$4.92)& $\mathbf{80.14(\pm3.32)}$ \\
   & Recall ($\%$) & 69.45($\pm$5.32) & 74.14($\pm$5.10)& 63.24($\pm$4.12) & 69.81($\pm$3.92)& 62.32($\pm$4.83)& 55.82($\pm$5.43)& $\mathbf{79.41(\pm4.57)}$\\
  \hline
  \multirow{3}{*}{D$-$N}& ACC($\%$) & 66.24($\pm$6.12) & 70.24($\pm$5.12)& 63.24($\pm$5.34) & 77.15($\pm$10.11)& 70.26($\pm$4.42)& 63.51($\pm$5.32)& $\mathbf{87.18(\pm3.12)}$\\
   & $ F_{1}$ ($\%$) & 65.14($\pm$5.42) & 69.82($\pm$4.42)& 64.27($\pm$5.47) & 78.13($\pm$8.12)& 71.24($\pm$9.32)& 61.35($\pm$7.17)& $ \mathbf{88.24(\pm4.11)}$ \\
   & Recall ($\%$) & 64.14($\pm$5.42) & 71.24($\pm$5.54)& 63.41($\pm$3.95) & 71.14($\pm$6.12)& 69.27($\pm$7.42)& 62.24($\pm$9.19)& $\mathbf{85.24(\pm4.17)}$\\
  \hline
  \multirow{3}{*}{F$-$H}& ACC($\%$) & 77.25($\pm$4.13) & 65.14($\pm$4.19)& 65.26($\pm$4.82) & 70.72($\pm$8.11)& 63.17($\pm$5.12)& 60.24($\pm$4.53)& $\mathbf{90.09(\pm4.12)}$ \\
   & $F_{1}$ ($\%$) & 75.52($\pm$3.31) & 64.49($\pm$5.12) & 63.83($\pm$4.19) & 70.29($\pm$5.43) & 61.08($\pm$5.62) & 59.04 ($\pm$5.12)& $\mathbf{91.30 (\pm4.34)}$ \\
   & Recall ($\%$) & 76.32($\pm$3.46) & 67.37($\pm$5.42) & 66.28($\pm$5.62) & 69.28($\pm$5.43) & 62.18($\pm$4.92) & 60.39($\pm$4.76) & $\mathbf{89.19(\pm4.09) }$ \\
  \hline
  \multirow{3}{*}{F$-$N}& ACC($\%$) & 68.23($\pm$4.42) & 66.21($\pm$4.52) & 60.45($\pm$5.02) & 70.27($\pm$7.19) & 60.45($\pm$5.12) & 57.19($\pm$8.17) & $\mathbf{89.10(\pm3.14) }$\\
   & $F_{1}$ ($\%$) & 66.42($\pm$4.17) & 60.32($\pm$4.12) & 60.42($\pm$4.31) & 68.17($\pm$5.12) & 61.26($\pm$6.12) & 58.31($\pm$4.92) & $\mathbf{88.26 (\pm3.63)}$ \\
   & Recall ($\%$) & 67.42($\pm$2.12) & 63.15($\pm$2.53) & 59.41($\pm$3.19)  & 70.53($\pm$4.12) & 59.21 ($\pm$4.12)& 59.27($\pm$3.16) & $\mathbf{90.19(\pm3.45) }$\\
  \hline
  \multirow{3}{*}{H$-$N}& ACC($\%$) & 55.14($\pm$3.12) & 60.28($\pm$3.53) & 64.32($\pm$3.19) & 61.28($\pm$8.17) & 63.32($\pm$4.12) & 61.21($\pm$3.15) & $\mathbf{95.19(\pm4.12)}$\\
   & $F_{1}$ ($\%$) & 54.32($\pm$3.12)  & 54.28($\pm$3.62) & 55.28($\pm$3.12) & 64.18($\pm$3.41) & 61.19($\pm$4.12) & 61.31($\pm$3.16) & $\mathbf{94.16(\pm3.11)}$ \\
   & Recall ($\%$) & 56.23($\pm$3.42) & 56.23($\pm$3.19) & 56.16($\pm$4.02) & 63.27($\pm$4.05) & 63.18($\pm$4.17) & 60.24($\pm$4.32) & $\mathbf{96.33(\pm3.12) }$\\
  \hline
  \multirow{3}{*}{Four}& ACC($\%$) & 33.42($\pm$5.12) & 31.18($\pm$5.31) & 43.27($\pm$3.17) & 35.29($\pm$3.42) & 37.31($\pm$8.12) & 30.41($\pm$5.32) & $\mathbf{65.19 (\pm3.96)}$\\
   & $F_{1}$ ($\%$) & 34.32($\pm$4.52) & 35.26($\pm$5.19) & 36.28 ($\pm$4.98)& 33.52($\pm$4.44) & 31.48($\pm$7.12) & 33.32($\pm$4.12) & $\mathbf{65.43(\pm3.19)}$ \\
   & Recall($\%$) & 35.53($\pm$6.17) & 30.36($\pm$5.11) & 31.32($\pm$4.12)  & 33.37($\pm$3.13) & 43.42($\pm$4.92) & 32.28 ($\pm$6.11)& $\mathbf{63.41(\pm3.34)}$\\
  \hline
  \end{tabular}
\end{table*}

 \subsubsection{The results of comparision decoding model}

At present, there are no large-scale public data sets in the field of neural decoding, and most of the decoding models reported are mainly concentrated on specific tasks or data sets. These methods usually carry out some special processing or adopt some prior knowledge to constraint, which will lead to difficulties in horizontal comparison of the performance of various decoding methods.  This article summarized the performance of the neural decoding models reported in recent years in the table \ref{tab:6}.

\begin{table*}[htbp]
  \centering
  \caption{The results of the latest neural decoding model}\label{tab:6}
  \begin{tabular}{cccccc}
  \toprule
  Model& Model(yesrs)& data sets & Categories number & Accuracy($\%$) & Reference\\
  \midrule
  \multirow{5}{*}{Traditional model}&Linear SVM $+$ network pattern(2019)&EEG&2&95.34& \cite{rezvani2019} \\
                           &RBF SVM $+$ network pattern(2019)&EEG&2&92.00& ~\cite{al2020decoding}\\
                         &Linear SVM $+$ network pattern(2020)&EEG&2&86.67& ~\cite{sun2020emotion} \\
                         &SVM $+$ network pattern(2020)&EEG&2&89.50& ~\cite{Geravanchizadeh_2020} \\
                         &RF $+$ network pattern(2020)&MEG&2&78.56& ~\cite{Yang2020The}\\
  \hline

  \multirow{4}{*}{Tensor model} &M2E $+$ network pattern(2018)&fMRI&2&71.43($\pm$1.00)& ~\cite{liu2018multi} \\
                          &HOSVD $+$ network pattern(2020)&fMRI&2&80.43& ~\cite{noroozi2020tensor}  \\
                         &MSTD $+$ network pattern(2020)&fMRI&2&85.91&~\cite{zhang2020multi}  \\
                          &ALS $+$ network pattern(2021)&fMRI&2&94.00& ~\cite{9146335} \\
  \hline
  \multirow{6}{*}{Deep learning Model} &LSTM $+$ network pattern(2019)&fMRI&2&90.60($\pm$2.70)& ~\cite{LI2019116059}  \\
                               & EEGNet(2020)&EEG&2&78.46($\pm$12.50) & ~\cite{9201053}\\
                        & CNN $+$ network pattern(2020)&EEG&2&80.74($\pm$1.48)& ~\cite{li2020deep}\\
                        & BiLSTM$+$ network pattern(2021)&EEG&4&71.05&~\cite{kwon2021visual} \\
                         & LSTM$+$ dynamic network pattern(2021)&EEG&2&85.32($\pm$9.09)&~\cite{Shamsi2021}  \\
  \bottomrule
  \end{tabular}
\end{table*}

Seen from the results of Table \ref{tab:4} and Table \ref{tab:6}, For the fMRI data set, the highest decoding rate of the STN model about two classification tasks is $95.16\%$, the highest four classification decoding is $64.32\%$. For EEG/MEG data, the highest decoding rate of the STN model about two classification tasks is $96.79\%$. Therefore, combining with the comparison results of Table \ref{tab1:5} on three type of decoding models, we found that the method in this paper, that is STN model, its decoding performances are exceed the decoding performance of most models. These comparison results also show that the decoding effect of the STN model proposed in this paper is certainly dominant.

\subsection{Analysis of model parameters}

For the parameters of the STN model, this article mainly analyzed some of these parameters, and one was $k$, which was the number of feature dimensions, and the other were $\alpha$ and $\beta$, which control visual stimulus category features and constraint brain network model information , last was $\gamma$, which was the hyperparameter of the constraint matrix $W^{(s)}$. In order to analyze the influence of specified parameters on the model, this paper adopts the way of fixing other parameters, and adjust the target parameters, then observing their influence on the decoding rate of the STN model.

\begin{figure}[!ht]
  \centering
  \includegraphics[width=0.80\linewidth]{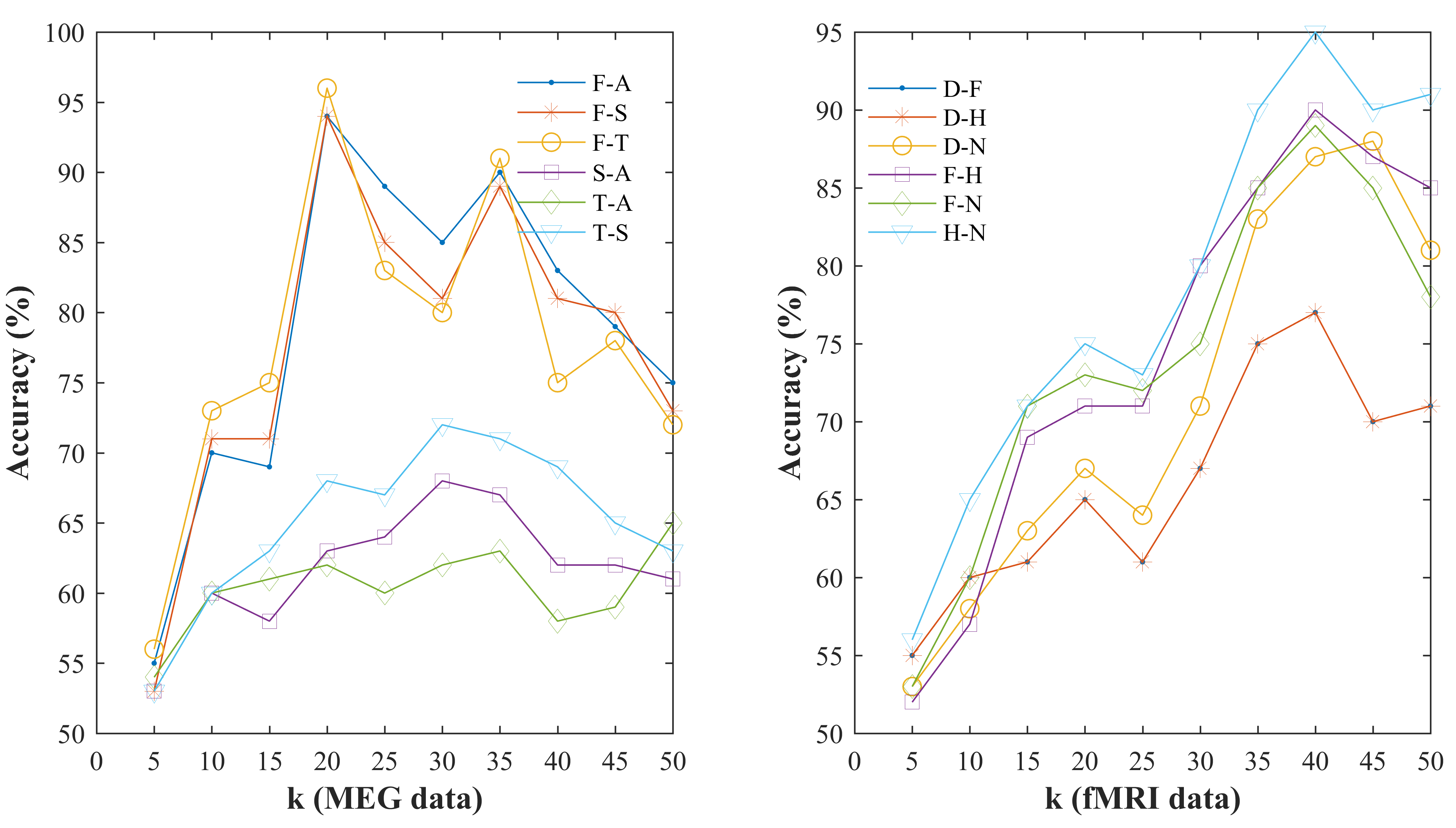}
  \caption{The change of the decoding rate about model for different parameters $k$ on two modal data sets}\label{fig:5.4}
\end{figure}

The hyperparameter $k$ was the number of components in the stimulus modality of the tensor decomposition process. It was also the number of feature dimension extracted from the coupled space between brain network pattern features and visual stimuli features. To analyze the influence of parameters, this paper selected the range to analyze the change of decoding rate. The range is from 0 to 50 components with a step size of 5.

AS seen from the decoding changes present in figure \ref{fig:5.4}, when the value of $k$ is small, model's decoding rate is not ideal on two modalities data. However, when the values of $k$ is set about 20, the STN model shows a better decoding effect on MEG data sets. For fMRI data sets, the STN model shows a better decoding effect when the value of $k$ is about 40.

\begin{figure}[!ht]
  \centering
  \includegraphics[width=0.80\linewidth]{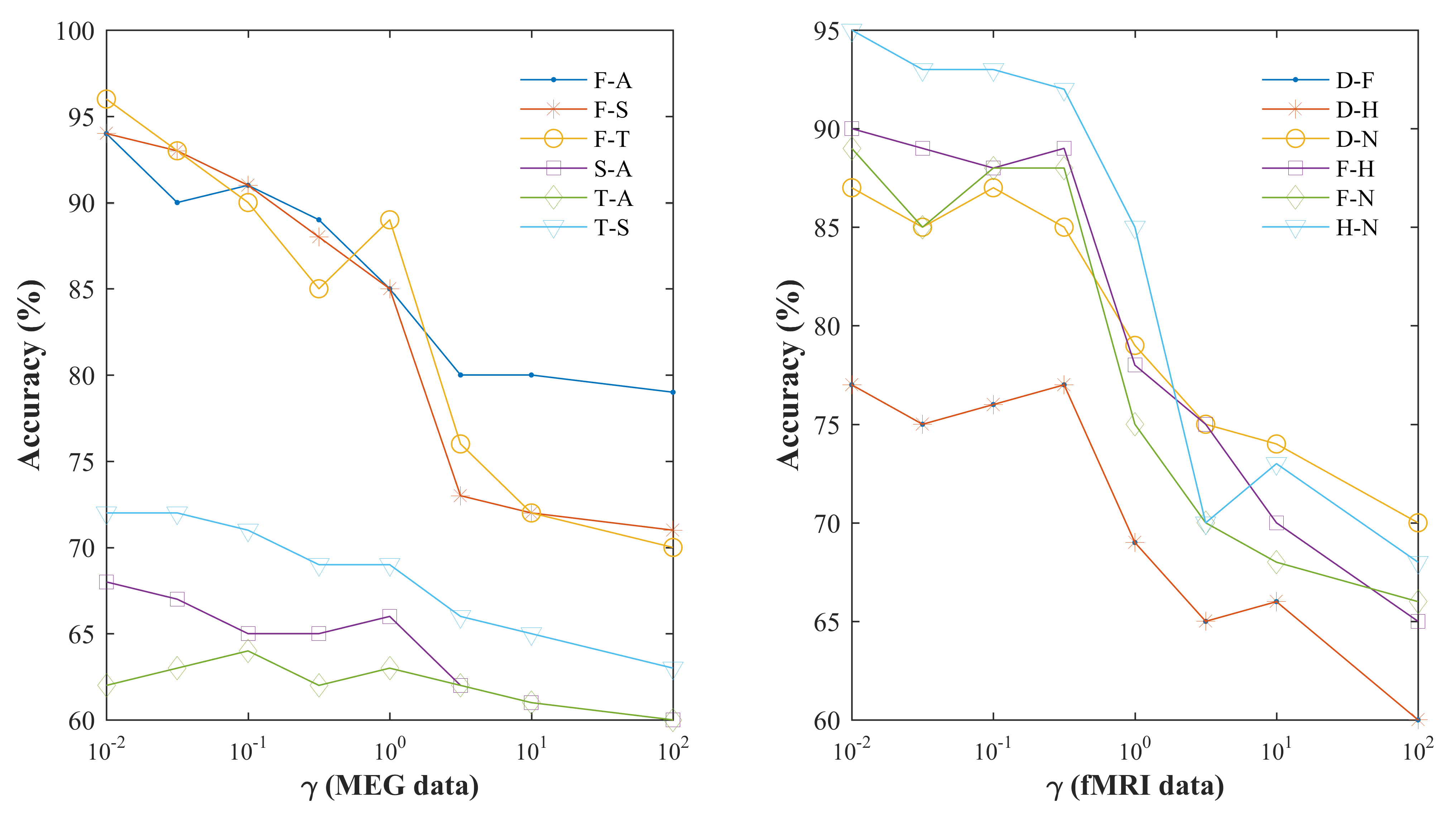}
  \caption{The change of the decoding rate about model for different parameters $\gamma$ on two modal data sets}\label{fig:5.5}
\end{figure}

The hyperparameter $\gamma$ was the regularization parameter of the variable $W^{(s)}$ in the STN model, and it was used to constrain the mapping relationship from the brain network feature space to the visual stimulus space. The paper sets the adjustment range of the parameter $\gamma$ to be $\{10^ {-2},5\times10^{-2},\cdots,10^{2}\}$ , and then analyzes the influence on decoding the decoding rate of the model on multiple data sets. However, the results of Figure \ref{fig:5.5} show that the decoding rate change of the STN model is not very obvious on the different values of parameter $\gamma$.

\begin{figure}[!ht]
  \centering
  \includegraphics[width=0.8\linewidth]{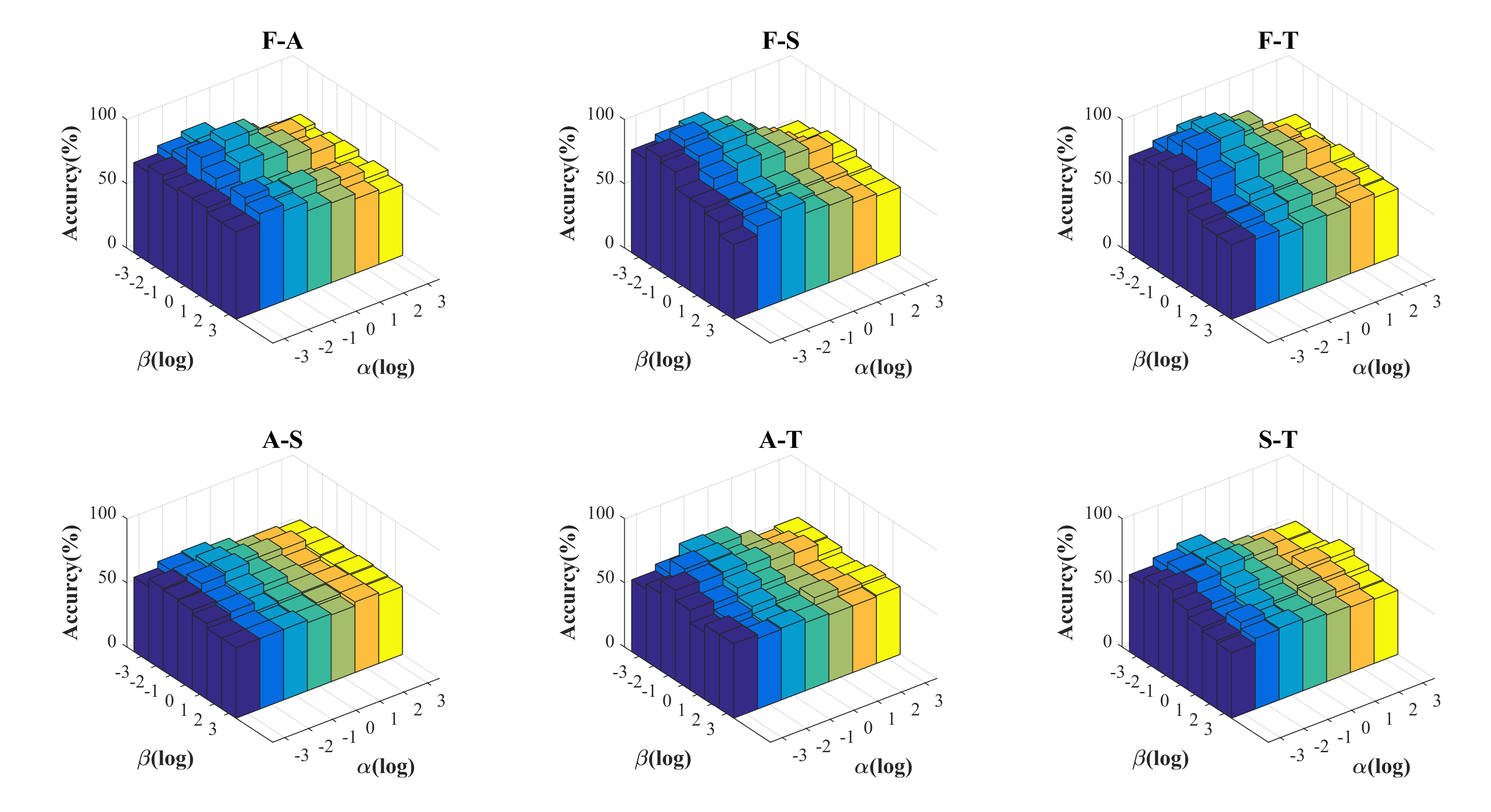}
  \caption{The change of the decoding rate about model for different parameters $\alpha$ and $\beta$ on MEG data sets}\label{fig:5.6}
\end{figure}

\begin{figure}[!ht]
  \centering
  \includegraphics[width=0.8\linewidth]{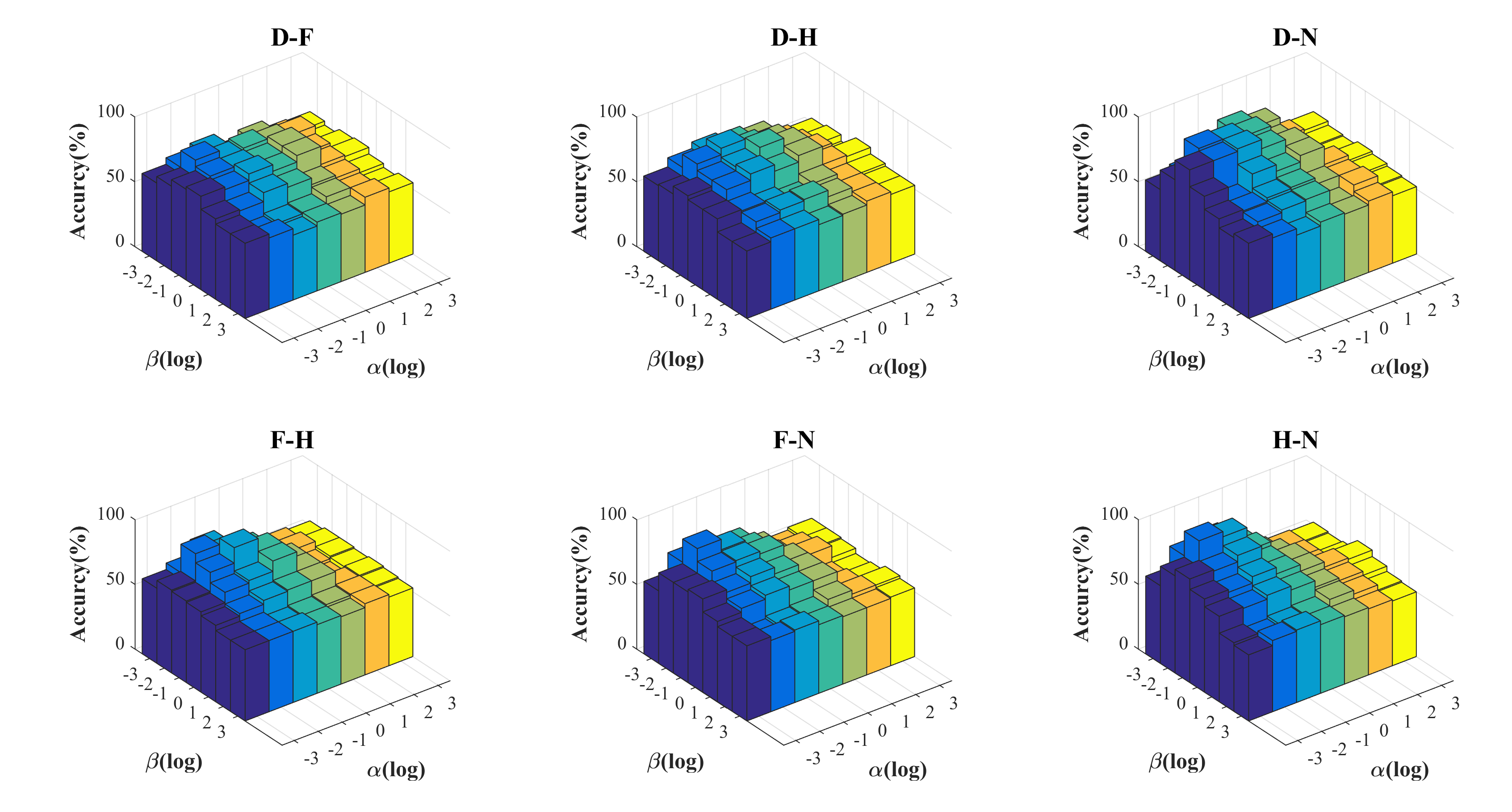}
  \caption{The change of the decoding rate about model for different parameters $\alpha$ and $\beta$ on MEG data sets}\label{fig:5.7}
\end{figure}

The hyperparameters $\alpha$ and $\beta$ were two parameters introduced by STN model, when the model solved the visual stimulus factor matrix. Where $\alpha$ control the similarity of brain network pattern under the visual stimulus, $\beta$ control the classification model between the visual category and the corresponding brain network pattern. These two parameters jointly constrain the model's variable $W^ {(s)}$ . Therefore, this paper put the adjustment process of these two parameters together, and the adjustment range is both $\{10^{-3},10^{-2},\cdots,10^{3}\}$, the decoding rate of the model are shown in figure \ref{fig:5.6} and \ref{fig:5.7}.

Seen from figure \ref{fig:5.6} and \ref{fig:5.7}, the ideal decoding results among STN model are both concentrated in the area of $\alpha,\beta\in(10^ {-2},10^{-1})\times(10^ {-2},10^{-1})$, whether it is on the MEG data or the fMRI data sets.

\section{Disscussion}

The paper proposed the STN model, which enables the decoding of neural activity patterns under different visual stimuli from  MEG and fMRI multiple data sets. The average decoding rate of binary classification about the STN model is $83.80\%$. In addition, four two-classification decoding rate are all higher than the average decoding rate of other models. The average of four-classification is 64.32$\%$, compared with the highest decoding accuracy of other models, its decoding ratio increased by $11.06\%$. In the decoding effect of two-classification data sets, the average decoding of the face and other three visual stimulus is generally higher than the decoding rate between the other two stimuli, and the average decoding rate of the face is $7.52\%$ better than the other two classifiers. These results indicate that the STN model is more likely to extract the characteristic pattern to distinguish the face and other visual stimuli. These results are also in line with some conclusions in neural decoding\cite{dima2018spatial}, that is, the brain activity pattern evoked by face stimuli is easier to distinguish. However, for the decoding effect between two visual stimuli, two decoding results are lower than other models, that may require subsequent improvements for the model to improve decoding accuracy.

On the fMRI data sets, the average decoding of two-classification about the STN model is $86.67\%$, which is higher by $18.46\%$ than the highest decoding rate of other models, and the average decoding of four-classification is $65.19\%$, that is higher by $21.92\%$ than the highest decoding rate of other models. Indeed, all the decoding rates are significantly higher than those of other models. These results show that the STN model has more substantial advantages in decoding emotion neural signal. This article speculates that the reason is that the tensor model, which is constrained by visual stimulus feature, is more able to extract graph structure features with discriminative semantics.

The STN model proposed in this research extracts the brain network characteristics by finding the best sub-graph pattern from the original graph structure. In recent years, the development of brain network is very rapid, which make it possible for many studies to use the whole brain network connectivity information to decode external stimuli and provide new ideas and directions for neural computing research\cite{manning2018a}. However, the complexity of brain network data and the lack of graph vector representation methods make it very difficult to directly introduce into the model to mine effective information. The most direct method is to extract graph-related features from the brain network structure, such as using the local weigh coefficient of ROI in the brain network\cite{2018Integrating}, topological core, and other features to study brain state\cite{shervashidze2011weisfeiler-lehman,morris2017glocalized}.However, this method is often poor in interpretability. The other is to extract the characteristics of the sub-graph pattern. In fact, the sub-graph pattern is more suitable for the brain network pattern. For example,it can model the network connnection pattern around the vertex and capture the changed in the local area\cite{kong2014Brain}. The sub-graph pattern feature can be in the form of a weighted graph\cite{2019Weighted,kong2013discriminative}. For example, Kong et al. proposed a probability distribution model based on dynamic programming, which scores the sub-graphs in each set of graph patterns\cite{kong2013discriminative}. The sub-graph can also be a sub-graph with a side view that introduced multiple vector constraints to find the optimal feature set of subgraphs for graph classification\cite{cao2016Mining}. This article used the second method, that is, from the perspective of visual stimulus features, to extract a set of sub-picture structure patterns and used them to classify brain network structure data. The article's experimental results prove this method's superiority in neural decoding.

In addition, this model in our paper added visual stimulus constraints to the tensor decomposition model. Currently, many studies have added constraint information to tensor decomposition to improve the effectiveness of the model. For example, Carroll\cite{carroll1980candelinc} et al. applied the linearly constrained least squares method to tensor data, Davidson\cite{davidson2013network} et al. proposed to apply alternately constrained least squares framework to analysis  brain network on fMRI data, and Wang\cite{Wang2015Rubik} et al. introduced knowledge-constrained tensor decomposition to calculate the representation analysis. These research results, except for the low-rank hypothesis, generally use the relationship between data or behavioral data as auxiliary variables to improve the quality of tensor decomposition. However, a problem faced by the current tensor calculation model is that some prior guidance and constraints are specifically designed for a specific field, resulting in these methods not working in other fields. Therefore, it is important to introduce prior knowledge to analysis tensor brain network. For example,in the STN model in our research, an important priori hypothesis is that if the characteristics of stimulus are similar, then the brain network patterns are similar. In the follow-up study of this model, we can add some constraints to further analyze the effect of decoding. For example, certain brain regions have been proven to be involved in the processing of a specific cognitive process, and then in order to preserve the adjacent areas around these brain regions and find other vertices at the same time, the known brain regions can be used as a mask to restrict the model, so that to better discover the brain areas that match the mask. For the subsequent models in this paper, we can put such as the FFA brain area, which is known to involve in the processing of human faces\cite{jonas2018a}, the amygdala, which is involved in emotion processing\cite{10.1093/cercor/bhaa176,PHAN2002331}, and other known brain areas into the model as local constraints.

\section{Conclusion}

The experimental results of this paper show that the STN model can extract the discrimination pattern of the brain network and achieve to decode the visual stimulus category. Combining the tensor decomposition, which was constrained with the prior visual information, this paper uses graph knowledge and tensor structure to represent the brain network pattern and construct the tensor brain network model. During STN models' learning, it first extracts discrimination sub-brain network pattern from the original brain network, then sets the sub-network pattern as a feature into the classification frame to decode visual stimulus category. The experimental results show that the sub-network pattern extracted from the STN model has a stronger ability to represent the brain activity pattern under different visual stimuli, significantly improving the accuracy of decoding neural signal, especially for emotional decoding with abstract semantics.


\bibliographystyle{IEEEtran}
\bibliography{tensornetwork}

\begin{thebibliography}{10}
\providecommand{\url}[1]{#1}
\csname url@samestyle\endcsname
\providecommand{\newblock}{\relax}
\providecommand{\bibinfo}[2]{#2}
\providecommand{\BIBentrySTDinterwordspacing}{\spaceskip=0pt\relax}
\providecommand{\BIBentryALTinterwordstretchfactor}{4}
\providecommand{\BIBentryALTinterwordspacing}{\spaceskip=\fontdimen2\font plus
\BIBentryALTinterwordstretchfactor\fontdimen3\font minus
  \fontdimen4\font\relax}
\providecommand{\BIBforeignlanguage}[2]{{%
\expandafter\ifx\csname l@#1\endcsname\relax
\typeout{** WARNING: IEEEtran.bst: No hyphenation pattern has been}%
\typeout{** loaded for the language `#1'. Using the pattern for}%
\typeout{** the default language instead.}%
\else
\language=\csname l@#1\endcsname
\fi
#2}}
\providecommand{\BIBdecl}{\relax}
\BIBdecl

\bibitem{kriegeskorte2019interpreting}
N.~Kriegeskorte and P.~K. Douglas, ``Interpreting encoding and decoding
  models,'' \emph{Current Opinion in Neurobiology}, vol.~55, pp. 167--179,
  2019.

\bibitem{2011Decoding}
A.~M. Chan, E.~Halgren, K.~Marinkovic, and S.~S. Cash, ``Decoding word and
  category specific spatiotemporal representations from meg and eeg,''
  \emph{NeuroImage}, vol.~54, no.~4, pp. 3028--3039, 2011.

\bibitem{2020Decoding}
R.~Al-Fahad, M.~Yeasin, and G.~M. Bidelman, ``Decoding of single-trial eeg
  reveals unique states of functional brain connectivity that drive rapid
  speech categorization decisions,'' \emph{Journal of Neural Engineering},
  vol.~17, no.~1, p. 016045 (18pp), 2020.

\bibitem{Anzellotti2018Beyond}
S.~Anzellotti and M.~N. Coutanche, ``Beyond functional connectivity:
  Investigating networks of multivariate representations,'' \emph{Trends in
  Cognitive Sciences}, vol.~22, no.~3, pp. 258--269, 2018.

\bibitem{fatemeh2019statistical}
Z.~J. Fatemeh, B.~Atena, and R.~D. Mohammad, ``Statistical algorithms for
  emotion classification via functional connectivity,'' \emph{Journal of
  Integrative Neuroscience}, vol.~18, no.~3, pp. 293--297, 2019.

\bibitem{wang2019decoding}
X.~Wang, J.~Gu, J.~Xu, X.~Li, J.~Geng, B.~Wang, and B.~Liu, ``Decoding natural
  scenes based on sounds of objects within scenes using multivariate pattern
  analysis,'' \emph{Neuroscience Research}, vol. 148, pp. 9--18, 2019.

\bibitem{parhizi2018decoding}
B.~Parhizi, M.~R. Daliri, and M.~Behroozi, ``Decoding the different states of
  visual attention using functional and effective connectivity features in fmri
  data,'' \emph{Cognitive Neurodynamics}, vol.~12, no.~2, pp. 157--170, 2018.

\bibitem{gonzalezcastillo2015tracking}
J.~Gonzalezcastillo, C.~W. Hoy, D.~A. Handwerker, M.~E. Robinson, L.~C.
  Buchanan, Z.~S. Saad, and P.~A. Bandettini, ``Tracking ongoing cognition in
  individuals using brief, whole-brain functional connectivity patterns,''
  \emph{Proceedings of the National Academy of Sciences of the United States of
  America}, vol. 112, no.~28, pp. 8762--8767, 2015.

\bibitem{cao2015a}
B.~Cao, X.~Kong, and P.~S. Yu, ``A review of heterogeneous data mining for
  brain disorder identification,'' \emph{Brain Informatics}, vol.~2, no.~4, pp.
  253--264, 2015.

\bibitem{huang2019tensor}
S.~Huang, H.~Peng, Y.~Chen, K.~Sun, F.~Shen, T.~Wang, and T.~Ma, ``Tensor
  discriminant analysis for mi-eeg signal classification using convolutional
  neural network,'' in \emph{Annual International Conference of the IEEE
  Engineering in Medicine and Biology Society. IEEE Engineering in Medicine and
  Biology Society}, 2019, pp. 5971--5974.

\bibitem{2017Detecting}
S.~Aviyente, E.~Al-Sharoa, and M.~Al-Khassaweneh, ``Detecting brain dynamics
  during resting state: a tensor based evolutionary clustering approach,'' in
  \emph{Wavelets and Sparsity XVII}, 2017.

\bibitem{ZHU2020116924}
``Discovering dynamic task-modulated functional networks with specific spectral
  modes using meg,'' \emph{NeuroImage}, vol. 218, p. 116924, 2020.

\bibitem{alsharoa2019tensor}
E.~Alsharoa, M.~Alkhassaweneh, and S.~Aviyente, ``Tensor based temporal and
  multilayer community detection for studying brain dynamics during resting
  state fmri,'' \emph{IEEE Transactions on Biomedical Engineering}, vol.~66,
  no.~3, pp. 695--709, 2019.

\bibitem{Topology2020}
Y.~Shen, X.~Fu, G.~B. Giannakis, and N.~D. Sidiropoulos, ``Topology
  identification of directed graphs via joint diagonalization of correlation
  matrices,'' \emph{IEEE Transactions on Signal and Information Processing over
  Networks}, vol.~6, pp. 271--283, 2020.

\bibitem{Cao2017T-BNE}
B.~Cao, L.~He, X.~Wei, M.~Xing, P.~S. Yu, H.~Klumpp, and A.~D. Leow, ``T-bne:
  Tensor-based brain network embedding,'' in \emph{the 2017 SIAM International
  Conference on Data Mining}, 2017, pp. 189--197.

\bibitem{2018Tensor}
R.~Mayhugh, C.~Hugenschmidt, J.~Rejeski, P.~Laurienti, and F.~Mokhtari,
  ``Tensor-based vs. matrix-based rank reduction in dynamic brain
  connectivity,'' in \emph{Image Processing}, 2018.

\bibitem{cowen2014neural}
A.~S. Cowen, M.~M. Chun, and B.~A. Kuhl, ``Neural portraits of perception:
  Reconstructing face images from evoked brain activity,'' \emph{NeuroImage},
  vol.~94, pp. 12--22, 2014.

\bibitem{narita2011Tensor}
A.~Narita, K.~Hayashi, R.~Tomioka, and H.~Kashima, ``Tensor factorization using
  auxiliary information,'' in \emph{Joint European Conference on Machine
  Learning and Knowledge Discovery in Databases}, 2011.

\bibitem{Eckstein1992On}
J.~Eckstein and D.~P. Bertsekas, ``On the douglas-rachford splitting method and
  the proximal point algorithm for maximal monotone operators,''
  \emph{Mathematical Programming}, vol.~55, no.~1, pp. 293--318, 1992.

\bibitem{lin2011Linearized}
Z.~Lin, R.~Liu, and Z.~Su, ``Linearized alternating direction method with
  adaptive penalty for low-rank representation,'' \emph{Advances in Neural
  Information Processing Systems}, pp. 612--620, 2011.

\bibitem{gao2019parallelizable}
B.~Gao, X.~Liu, and Y.~Yuan, ``Parallelizable algorithms for optimization
  problems with orthogonality constraints,'' \emph{SIAM Journal on Scientific
  Computing}, vol.~41, no.~3, pp. 1949--1983, 2019.

\bibitem{OptimizationAlgorithmsonMatrixManifolds}
P.-A. Absil, R.~Mahony, and R.~Sepulchre, \emph{Optimization Algorithms on
  Matrix Manifolds}.\hskip 1em plus 0.5em minus 0.4em\relax Princeton:
  Princeton University Press, 11 Apr. 2009.

\bibitem{10.1007/s10107-012-0584-1}
Z.~Wen and W.~Yin, ``A feasible method for optimization with orthogonality
  constraints,'' vol. 142, no. 1–2, 2013.

\bibitem{2019Network}
Y.~Liang, B.~Liu, J.~Ji, and X.~Li, ``Network representations of facial and
  bodily expressions: Evidence from multivariate connectivity pattern
  classification,'' \emph{Frontiers in Neuroscience}, vol.~13, p. 1111, 2019.

\bibitem{ajmera2020decoding}
S.~Ajmera, H.~Jain, M.~Sundaresan, and D.~Sridharan, ``Decoding task-specific
  cognitive states with slow, directed functional networks in the human
  brain,'' \emph{eNeuro}, vol.~7, no.~4, pp. 0512--19, 2020.

\bibitem{tottenham2009the}
N.~Tottenham, J.~W. Tanaka, A.~C. Leon, T.~Mccarry, M.~Nurse, T.~A. Hare, D.~J.
  Marcus, A.~Westerlund, B.~J. Casey, and C.~A. Nelson, ``The nimstim set of
  facial expressions: Judgments from untrained research participants,''
  \emph{Psychiatry Research-neuroimaging}, vol. 168, no.~3, pp. 242--249, 2009.

\bibitem{dailey2001california}
M.~Dailey, G.~Cottrell, and J.~Reilly, ``California facial expressions, cafe,
  unpublished digital images,'' \emph{Computer Science and Engineering
  Department, UCSD}, 2001.

\bibitem{lyons1998coding}
M.~J. Lyons, S.~Akamatsu, M.~Kamachi, and J.~Gyoba, ``Coding facial expressions
  with gabor wavelets,'' in \emph{Third IEEE International Conference on
  Automatic Face and Gesture Recognition}, 1998, pp. 200--205.

\bibitem{langner2010presentation}
O.~Langner, R.~Dotsch, G.~Bijlstra, D.~H.~J. Wigboldus, S.~T. Hawk, and
  A.~Van~Knippenberg, ``Presentation and validation of the radboud faces
  database,'' \emph{Cognition Emotion}, vol.~24, no.~8, pp. 1377--1388, 2010.

\bibitem{liu2020rapidly}
C.~Liu, Y.~Kang, L.~Zhang, and J.~Zhang, ``Rapidly decoding image categories
  from meg data using a multivariate short-time fc pattern analysis approach,''
  \emph{IEEE Journal of Biomedical and Health Informatics}, vol.~25, no.~4, pp.
  1139--1150, 2020.

\bibitem{rezvani2019}
Z.~Rezvani, M.~Zare, G.~{\v{Z}}ari{\'c}, M.~Bonte, J.~Tijms, M.~Van~der Molen,
  and G.~F. Gonz{\'a}lez, ``Machine learning classification of dyslexic
  children based on eeg local network features,'' \emph{BioRxiv}, p. 569996,
  2019.

\bibitem{Yang2020The}
T.~Yang, K.~Zhao, and T.~Chen, ``The predominant functional connections of
  recognizing fear and surprise expression: a meg study,'' in \emph{2020 5th
  International Conference on Intelligent Informatics and Biomedical Sciences
  (ICIIBMS)}, 2020, pp. 65--69.

\bibitem{sharan2017}
V.~Sharan and G.~Valiant, ``Orthogonalized als: A theoretically principled
  tensor decomposition algorithm for practical use,'' in \emph{International
  Conference on Machine Learning}.\hskip 1em plus 0.5em minus 0.4em\relax PMLR,
  2017, pp. 3095--3104.

\bibitem{Yin2018Joint}
K.~Yin, W.~K. Cheung, Y.~Liu, B.~C.~M. Fung, and J.~Poon, ``Joint learning of
  phenotypes and diagnosis-medication correspondence via hidden interaction
  tensor factorization,'' in \emph{Twenty-Seventh International Joint
  Conference on Artificial Intelligence IJCAI-18}, 2018.

\bibitem{Yin2019Learning}
K.~Yin, D.~Qian, W.~K. Cheung, B.~C.~M. Fung, and J.~Poon, ``Learning
  phenotypes and dynamic patient representations via rnn regularized collective
  non-negative tensor factorization,'' in \emph{the AAAI Conference on
  Artificial Intelligence}, vol.~33, 2019, pp. 1246--1253.

\bibitem{Comon2009tensor}
P.~Comon, X.~Luciani, and A.~L. De~Almeida, ``Tensor decompositions,
  alternating least squares and other tales,'' \emph{Journal of Chemometrics},
  vol.~23, no. 7‐8, pp. 393--405.

\bibitem{Wang2015Rubik}
Y.~Wang, R.~Chen, J.~Ghosh, J.~C. Denny, A.~Kho, Y.~Chen, B.~A. Malin, and
  J.~Sun, ``Rubik: Knowledge guided tensor factorization and completion for
  health data analytics.''\hskip 1em plus 0.5em minus 0.4em\relax New York, NY,
  USA: Association for Computing Machinery, 2015.

\bibitem{Hochreiter1997long}
S.~Hochreiter and J.~Schmidhuber, ``Long short-term memory,'' \emph{Neural
  Computation}, vol.~9, no.~8, pp. 1735--1780, 1997.

\bibitem{Lawhern2018EEG}
V.~J. Lawhern, A.~J. Solon, N.~R. Waytowich, S.~M. Gordon, C.~P. Hung, and
  B.~J. Lance, ``\{EEGNet\}$:$ a compact convolutional neural network for
  \{EEG\}-based brain{\textendash}computer interfaces,'' \emph{Journal of
  Neural Engineering}, vol.~15, no.~5, p. 056013, jul 2018.

\bibitem{Destrieux2010Automatic}
C.~Destrieux, B.~Fischl, A.~Dale, and E.~Halgren, ``Automatic parcellation of
  human cortical gyri and sulci using standard anatomical nomenclature,''
  \emph{NeuroImage}, vol.~53, no.~1, pp. 1--15, 2010.

\bibitem{yan2016dpabi}
C.~Yan, X.~Wang, X.~Zuo, and Y.~Zang, ``Dpabi: Data processing $\&$ analysis
  for (resting-state) brain imaging,'' \emph{Neuroinformatics}, vol.~14, no.~3,
  pp. 339--351, 2016.

\bibitem{al2020decoding}
R.~Al-Fahad, M.~Yeasin, and G.~M. Bidelman, ``Decoding of single-trial eeg
  reveals unique states of functional brain connectivity that drive rapid
  speech categorization decisions,'' \emph{Journal of Neural Engineering},
  vol.~17, no.~1, p. 016045, 2020.

\bibitem{sun2020emotion}
X.~Sun, B.~Hu, X.~Zheng, Y.~Yin, and C.~Ji, ``Emotion classification based on
  brain functional connectivity network,'' in \emph{2020 IEEE International
  Conference on Bioinformatics and Biomedicine (BIBM)}.\hskip 1em plus 0.5em
  minus 0.4em\relax IEEE, 2020, pp. 2082--2089.

\bibitem{Geravanchizadeh_2020}
M.~Geravanchizadeh and S.~B. Gavgani, ``Selective auditory attention detection
  based on effective connectivity by single-trial eeg,'' \emph{Journal of
  Neural Engineering}, vol.~17, no.~2, p. 026021, 2020.

\bibitem{liu2018multi}
Y.~Liu, L.~He, B.~Cao, P.~Yu, A.~Ragin, and A.~Leow, ``Multi-view multi-graph
  embedding for brain network clustering analysis,'' in \emph{the AAAI
  Conference on Artificial Intelligence}, vol.~32, no.~1, 2018.

\bibitem{noroozi2020tensor}
A.~Noroozi and M.~Rezghi, ``A tensor-based framework for rs-fmri classification
  and functional connectivity construction,'' \emph{Frontiers in
  Neuroinformatics}, vol.~14, 2020.

\bibitem{zhang2020multi}
Y.-P. Zhang, L.~Xiao, G.~Zhang, B.~Cai, J.~M. Stephen, T.~W. Wilson, V.~D.
  Calhoun, and Y.-P. Wang, ``Multi-paradigm fmri fusion via sparse tensor
  decomposition in brain functional connectivity study,'' \emph{IEEE Journal of
  Biomedical and Health Informatics}, 2020.

\bibitem{9146335}
P.~B. Gender and I.~from fMRI~via Dynamic Functional~Connectivity, ``Predicting
  biological gender and intelligence from fmri via dynamic functional
  connectivity,'' \emph{IEEE Transactions on Biomedical Engineering}, vol.~68,
  no.~3, pp. 815--825, 2021.

\bibitem{LI2019116059}
``Interpretable, highly accurate brain decoding of subtly distinct brain states
  from functional mri using intrinsic functional networks and long short-term
  memory recurrent neural networks,'' \emph{NeuroImage}, vol. 202, p. 116059,
  2019.

\bibitem{9201053}
A.~David and H.~Mitsuhiro, ``Decoding hand motor imagery tasks within the same
  limb from eeg signals using deep learning,'' \emph{IEEE Transactions on
  Medical Robotics and Bionics}, vol.~2, no.~4, pp. 692--699, 2020.

\bibitem{li2020deep}
Y.~Li, J.~Liu, Z.~Tang, and B.~Lei, ``Deep spatial-temporal feature fusion from
  adaptive dynamic functional connectivity for mci identification,'' \emph{IEEE
  Transactions on Medical Imaging}, vol.~39, no.~9, pp. 2818--2830, 2020.

\bibitem{kwon2021visual}
B.-H. Kwon, J.-H. Jeong, and S.-W. Lee, ``Visual motion imagery classification
  with deep neural network based on functional connectivity,'' \emph{arXiv
  preprint arXiv:2103.02851}, 2021.

\bibitem{Shamsi2021}
S.~Foroogh, H.~Ali, and N.~Laleh, ``Early classification of motor tasks using
  dynamic functional connectivity graphs from {EEG},'' \emph{Journal of Neural
  Engineering}.

\bibitem{dima2018spatial}
D.~C. Dima, G.~Perry, and K.~D. Singh, ``Spatial frequency supports the
  emergence of categorical representations in visual cortex during natural
  scene perception,'' \emph{NeuroImage}, vol. 179, pp. 102--116, 2018.

\bibitem{manning2018a}
J.~R. Manning, X.~Zhu, T.~L. Willke, R.~Ranganath, K.~L. Stachenfeld,
  U.~Hasson, D.~M. Blei, and K.~A. Norman, ``A probabilistic approach to
  discovering dynamic full-brain functional connectivity patterns,''
  \emph{NeuroImage}, vol. 180, pp. 243--252, 2018.

\bibitem{2018Integrating}
X.~Cui, J.~Xiang, B.~Wang, J.~Xiao, Y.~Niu, and J.~Chen, ``Integrating the
  local property and topological structure in the minimum spanning tree brain
  functional network for classification of early mild cognitive impairment,''
  \emph{Frontiers in Neuroscience}, vol.~12, p. 701, 2018.

\bibitem{shervashidze2011weisfeiler-lehman}
N.~Shervashidze, P.~Schweitzer, E.~J. Van~Leeuwen, K.~Mehlhorn, and K.~M.
  Borgwardt, ``Weisfeiler-lehman graph kernels,'' \emph{Journal of Machine
  Learning Research}, vol.~12, pp. 2539--2561, 2011.

\bibitem{morris2017glocalized}
C.~Morris, K.~Kersting, and P.~Mutzel, ``Glocalized weisfeiler-lehman graph
  kernels: Global-local feature maps of graphs,'' in \emph{2017 IEEE
  International Conference on Data Mining (ICDM)}, 2017, pp. 327--336.

\bibitem{kong2014Brain}
X.~Kong and P.~S. Yu, ``Brain network analysis: a data mining perspective,''
  \emph{Acm Sigkdd Explorations Newsletter}, vol.~15, no.~2, pp. 30--38, 2014.

\bibitem{2019Weighted}
R.~Yu, L.~Qiao, M.~Chen, S.~W. Lee, X.~Fei, and D.~Shen, ``Weighted graph
  regularized sparse brain network construction for mci identification,''
  \emph{Pattern Recognition}, vol.~90, pp. 220--231, 2019.

\bibitem{kong2013discriminative}
X.~Kong, P.~S. Yu, X.~Wang, and A.~B. Ragin, ``Discriminative feature selection
  for uncertain graph classification,'' in \emph{the 2013 SIAM International
  Conference on Data Mining}, 2013, pp. 82--93.

\bibitem{cao2016Mining}
B.~Cao, X.~Kong, J.~Zhang, P.~S. Yu, and A.~B. Ragin, ``Mining brain networks
  using multiple side views for neurological disorder identification,'' in
  \emph{IEEE International Conference on Data Mining}, 2016.

\bibitem{carroll1980candelinc}
J.~D. Carroll, S.~Pruzansky, and J.~B. Kruskal, ``Candelinc: A general approach
  to multidimensional analysis of many-way arrays with linear constraints on
  parameters,'' \emph{Psychometrika}, vol.~45, no.~1, pp. 3--24, 1980.

\bibitem{davidson2013network}
I.~Davidson, S.~Gilpin, O.~Carmichael, and P.~B. Walker, ``Network discovery
  via constrained tensor analysis of fmri data,'' in \emph{the 19th ACM SIGKDD
  International Conference on Knowledge Discovery and Data Mining}, 2013, pp.
  194--202.

\bibitem{jonas2018a}
J.~Jonas, H.~Brissart, G.~Hossu, S.~Colnatcoulbois, J.~Vignal, B.~Rossion, and
  L.~Maillard, ``A face identity hallucination (palinopsia) generated by
  intracerebral stimulation of the face-selective right lateral fusiform
  cortex,'' \emph{Cortex}, vol.~99, pp. 296--310, 2018.

\bibitem{10.1093/cercor/bhaa176}
R.~M. Skiba and P.~Vuilleumier, ``{Brain Networks Processing Temporal
  Information in Dynamic Facial Expressions},'' \emph{Cerebral Cortex},
  vol.~30, no.~11, pp. 6021--6038, 06 2020.

\bibitem{PHAN2002331}
``Functional neuroanatomy of emotion: A meta-analysis of emotion activation
  studies in pet and fmri,'' \emph{NeuroImage}, vol.~16, no.~2, pp. 331 -- 348,
  2002.

\end{thebibliography}

\begin{IEEEbiography}[{\includegraphics[width = 1in,height = 1.25in, clip,keepaspectratio]{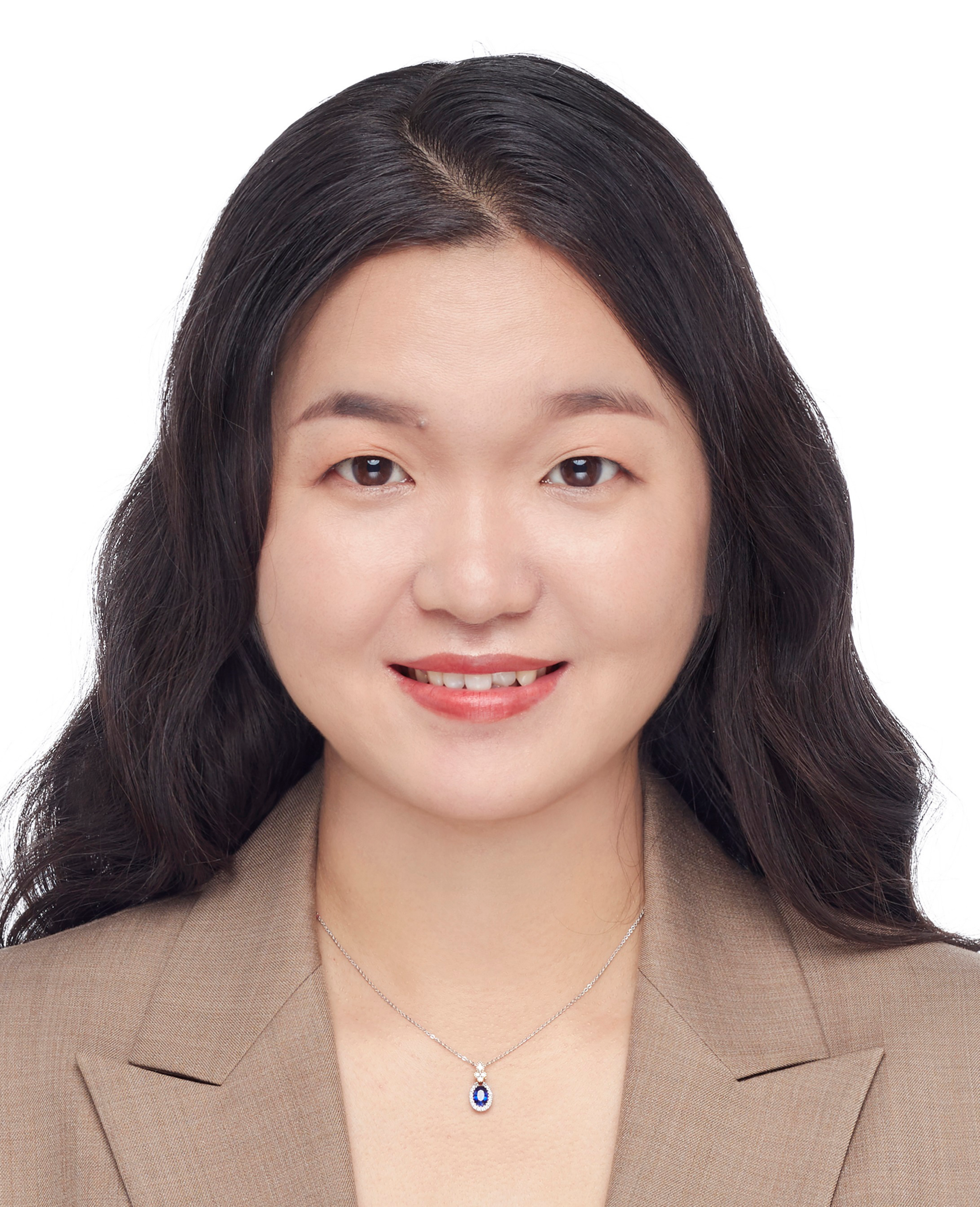}}]{Chunyu Liu}
received the B.S. degree in Mathematics and Applied Mathematics from Henan Normal University, the M.S. degree in Applied Mathematics from Northwest A$\&$F University, and the Ph.D. degree in Computer application technology from Beijing Normal University.

She is currently a Post-doctoral with the vision and brain imaging group in Peking University. Her currently research interests include
Machine learning, Neural decoding, and Visual attention.

\end{IEEEbiography}

\begin{IEEEbiography}[{\includegraphics[width = 1in,height = 1.25in, clip,keepaspectratio]{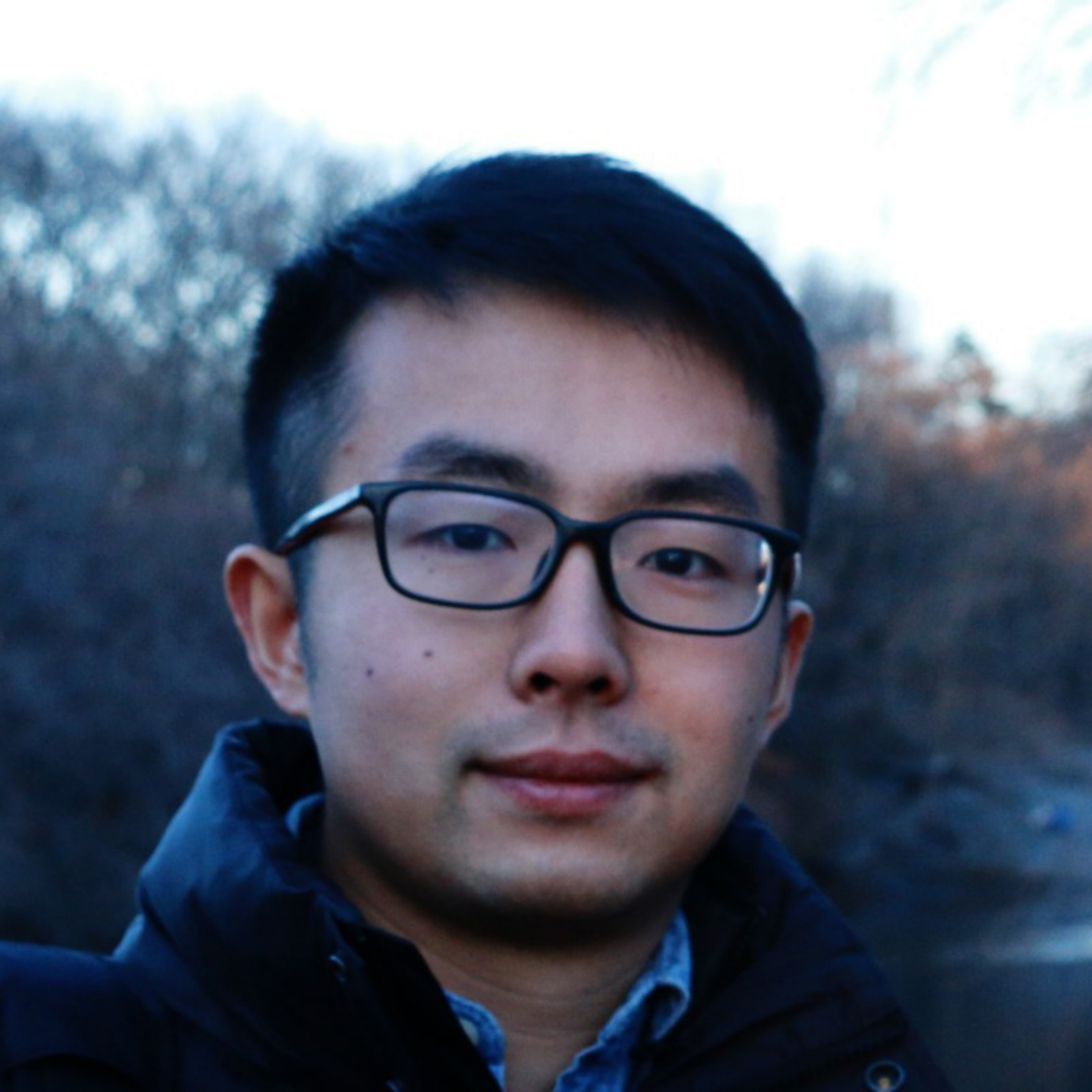}}]{Bokai Cao}

received Ph.D. in Computer Science from the University of Illinois at Chicago, B.Eng. in Computer Science and B.Sc. in Mathematics from Renmin University of China.

He is currently a research scientist at Meta Inc. where he focuses on user and video understanding with applications in modern recommendation systems.
\end{IEEEbiography}
\vspace{-180 mm}
\begin{IEEEbiography}[{\includegraphics[width = 1in,height = 1.25in, clip,keepaspectratio]{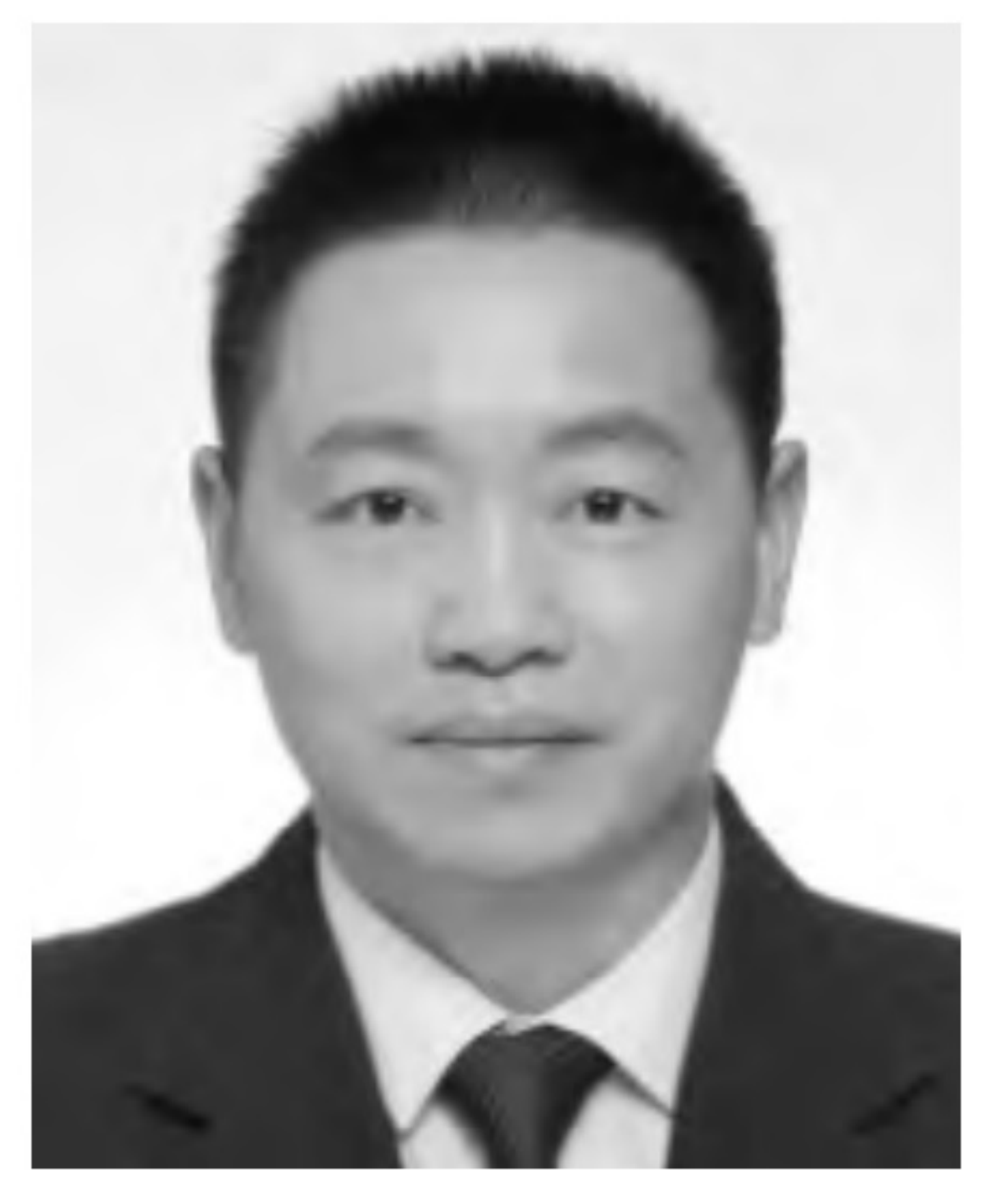}}]{Jiacai Zhang}
received the B.S. and M.S. degrees from Beijing Normal University, Beijing, China, in 2004, and the Ph.D. degree in pattern recognition and intelligent system from the Institute of Automation, Chinese Academy of Sciences, Beijing, China, in 2004.

He is currently a Professor with the College of Information Science and Technology, Beijing Normal University. He works on brain-computer interfaces and employs it to monitor and decode mental state continuously. His research enhances BCI systems with the ability to detect, process, respond to, and even regulate uses' brain activities, especially related to the affective state using physiological signals.
\end{IEEEbiography}

\end{document}